\documentclass[%
reprint,
amsmath,amssymb,
aps,
prb,
]{revtex4-1}

\usepackage{graphicx}
\usepackage{dcolumn}
\usepackage{bm}
\usepackage{hyperref}
\hypersetup{colorlinks,linkcolor={red},citecolor={blue},urlcolor={blue}}

\def\be{\begin{equation}}
\def\ee{\end{equation}}
\def\bea{\begin{eqnarray}}
\def\eea{\end{eqnarray}}
\def\v{{\bf v}}
\def\kv{{\bf k}}

\def\Sv{{\bf S}}

\def\sv{{\bm \sigma}}
\def\Rv{{\bf R}}
\def\rv{{\bf r}}
\def\no{\nonumber}
\def\ra{\rangle}
\def\la{\langle}
\def\s{\sigma}

\def\ve{\varepsilon}
\def\f{{\rm F}}
\def\T{\tilde {T}}
\def\esv{{\bf s}}

\usepackage{color}

\newcommand{\ket}[1]{\left|{#1}\right\rangle}

\begin{document}

\title{Transport in magnetically doped topological insulators:\\ Effects of magnetic clusters}

\author{A. N. Zarezad}
\affiliation{Department of Physics, Institute for Advanced Studies in Basic Sciences (IASBS), Zanjan 45137-66731, Iran}
\author{J. Abouie}
\email[]{jahan@iasbs.ac.ir}
\affiliation{Department of Physics, Institute for Advanced Studies in Basic Sciences (IASBS), Zanjan 45137-66731, Iran}

\begin{abstract}
We study the electron transport in a magnetically doped three
dimensional topological insulator (TI) by taking the effects of
impurity-impurity exchange interactions into account. The
interactions between magnetic impurities give rise to the formation
of {\it magnetic clusters} with temperature dependent mean sizes,
randomly distributed on the surface of the TI. Instead of dealing
with single magnetic impurities, we consider surface Dirac electrons
to be scattered off magnetic clusters, and define the scattering
potential in terms of clusters mean sizes. Within the semiclassical
Boltzmann approach, employing a generalized relaxation time
approximation, we obtain the surface conductivity of the TI by
solving four sets of recursive relations and demonstrate that, the
system is highly anisotropic and the surface conductivities possess
non-monotonic behaviors, they strongly depends on the direction, the
mean size and the number of magnetic clusters. We demonstrate that
the dependence of the anisotropic magnetoresistance (AMR) to the
spin direction of the magnetic clusters is inconsistent with the
angular dependence of the TI doped with non-interacting magnetic
impurities. Our results are consistent with the recent experiment on
the AMR of the Cr-doped $\rm {(Bi, Sb)}_2{\rm Te}_3$ TI.

\end{abstract}

\date{\today}


\maketitle


\section{Introduction}\label{sec:intro}
Topological insulators (TIs) are a new class of materials which
attracted intensive
theoretical\cite{PhysRevLett.95.146802,Science2006,PhysRevB.75.121306,PhysRevLett.98.106803}
and experimental\cite{Konig766, Hsieh2008} attentions in condensed
matter physics. TIs are band insulators in the bulk while conducting
along the surfaces, hosting metallic surface states with a
Dirac-cone-like dispersion. The Dirac cone is located in the center
of the Brillouin zone and the spin and momentum degrees of freedom
are locked. The spin-momentum locking of surface Dirac electrons is
protected by time-reversal
symmetry\cite{Science2006,Topological,PhysRevLett.95.226801,PhysRevLett.95.146802,PhysRevLett.96.106802,PhysRevB.76.045302},
which leads to a variety of interesting
effects\cite{PhysRevLett.102.216403, PhysRevB.78.195424, Yu61}, in
particular to the suppression of the elastic backscattering of
surface states in the absence of spin-flip
processes\cite{PhysRevB.82.155457,Culcer2}. The robustness of these
topologically protected processes, when introducing impurities which
break time-reversal symmetry, is of critical importance for
spin-based transport in such materials.

Recently, many efforts have been devoted to the investigation of the
surface transport properties of 3D TIs in the presence of localized
and identical magnetic impurities\cite{L.Andrew,Y.L.Chen2,X.F.Kou,
PhysRevB.83.245441}. In dilute magnetic TIs, the impurities are
uncorrelated and distributed randomly on the surface of the TI, thus
single-impurity-scattering processes (the scattering of the massless
Dirac electrons by a single magnetic impurity), specify the behavior
of the surface resistivity. In this case, the scattering amplitudes
and consequently the surface magneto-resistance strongly depend on
the orientation of impurities' spins\cite{Culcer2,
PhysRevLett.105.066802, 0953-8984-27-11-115301}. By increasing
doping the interactions of magnetic impurities become important and
their effects on the transport properties of the TI are crucial. The
multiple-impurity-scattering problem of the massless Dirac electrons
in the presence of magnetic impurities normal to the surface of the
TI has been evaluated theoretically in
Ref.\cite{PhysRevB.85.245433}. It has been shown that, different
from the single-impurity-scattering, the Hall component of
resistivity and the inverse momentum relaxation time exhibit
oscillatory behavior due to the interference during the double- and
triple-impurities-scattering processes. In spite of many studies on
the electron transport of magnetic TIs, the effects of magnetic
impurity interactions on the magneto-resistance of TIs have not been
addressed so far.

In this paper, we investigate the effects of magnetic
impurity-magnetic impurity exchange interactions on the transport in
a magnetic TI. We consider a 3D TI doped with magnetic impurities
localized on the surface of the TI, and investigate the surface
magneto-resistance of the TI by taking the exchange interactions of
magnetic impurities into account. The exchange interactions of
magnetic impurities give rise to the formation of "magnetic
clusters", constructed of correlated magnetic impurities with the
same direction. We consider that the surface massless Dirac
electrons scatter by magnetic clusters rather than single
impurities. The cluster concept was first introduced in Refs.
\cite{PhysRevB.84.024428, PhysRevB.77.165433} to explain the
temperature dependence of magneto-resistance of magnetic
semiconductors. The clusters' mean size (CMS) and the number of
clusters with mean sizes (CN) depend on temperature, they increase
by increasing temperature, pass through a maximum at a critical
temperature, and decrease toward zero at high
temperatures\cite{PhysRevB.84.024428, PhysRevB.77.165433}. We model
the scattering of massless Dirac electrons off magnetic clusters
with a potential, exponentially decaying with the electron-cluster
separation distance rescaled by the CMS. In order to obtain the
surface conductivity of such a system we employ the semi-classical
Boltzmann approach. However, since the scattering of Dirac electrons
by a magnetic cluster is not isotropic, the standard relaxation time
approximation is not applicable here. Employing a generalized
relaxation time approximation\cite{PhysRevB.79.045427} and solving
four sets of recursive relations we obtain the relaxation times of
Dirac electrons and demonstrate that, the system is highly
anisotropic and the surface conductivity strongly depends on the
CMS, the CN and the spin directions of magnetic clusters. We show
that surface conductivities possess a non-monotonic behavior, in
general they decrease sharply by increasing CMS and pass through a
minimum where the electrons Fermi wavelength is identically the same
as CMS. We demonstrate that the angular dependence of the AMR is
inconsistent with the angular dependence of the TI doped with
non-interacting magnetic impurities. We also show that our results
are consistent with the recent experiment on the AMR of the Cr-doped
$\rm {(Bi, Sb)}_2{\rm Te}_3$ TI.\cite{AMR-Exp}

The outline of this paper is as follows. In Sec. \ref{sec:model}, we
introduce our model and present the scattering potential in terms of
CMS. In Sec. \ref{sec:relaxation-time}, we obtain the transition
amplitudes of the system. Employing a generalized relaxation time
approximation we give a closed form for the effective relaxation
times of massless Dirac electrons. In Sec. \ref{sec:conductivity},
we compute the surface conductivities of the TI, both in isotropic
and anisotropic cases. In this section we investigate, in detail,
the behavior of the surface conductivities for different values of
CMS and clusters' spin directions. We also investigate the angular
dependence of the AMR for different values of CMS. In Sec.
\ref{sec:nonmagnetic-imp}, we also obtain the surface conductivity
of a 3D TI doped with non-magnetic impurities and compare the
behaviors of the relaxation times in both systems. We finally
summarize our results in Sec. \ref{sec:conclusion} and give the
concluding remarks.

\section{Our Model}\label{sec:model}

We consider a 3D topological insulator (TI) which its surface defines
the $xy$ plane and the surface is normal to $z$ direction. The minimal
Hamiltonian of the TI describing the dynamics of the surface
electrons, is given by \cite{JETLetters,Zhang}:
\begin{equation}
H_0=\hbar v_{\rm F}(\kv\times\sv)_z, \label{eq:TI-Hamiltonian}
\end{equation}
where $v_{\rm F}$ in the Fermi velocity, $\kv=(k_x,k_y)$ is the
vector of momentum and $\sv$ is the vector of Pauli matrices,
describing the electron spin. The eigenvalues and eigenvectors of
the Hamiltonian $H_0$ are given by:
\begin{equation}
\ve^{\pm}=\pm\hbar v_{\rm F}k, \label{eq:eigenenergy}
\end{equation}
and
\begin{equation}
\psi_{\kv,\pm}(\rv)=\frac{e^{i\kv\cdot\rv}}{\sqrt{2A}}\begin{pmatrix}
      \mp i {\rm e}^{-i \phi} \\
      1
\end{pmatrix},
\label{eq:eigenvector}
\end{equation}
where $\hbar$ is the Planck constant divided by $2\pi$,  $A$ is the TI surface area, $k=|\kv|=(k_x^2+k_y^2)^{1/2}$ is the modulus of the electron
momentum, and $\phi=\arctan(k_y/k_x)$ is the polar angle between
the momentum vector and the $k_x$ direction. It is important to note
that the spin states (\ref{eq:eigenvector}) are always perpendicular
to the motion direction.
\begin{figure}[ht]
\centerline{\includegraphics[width=90mm]{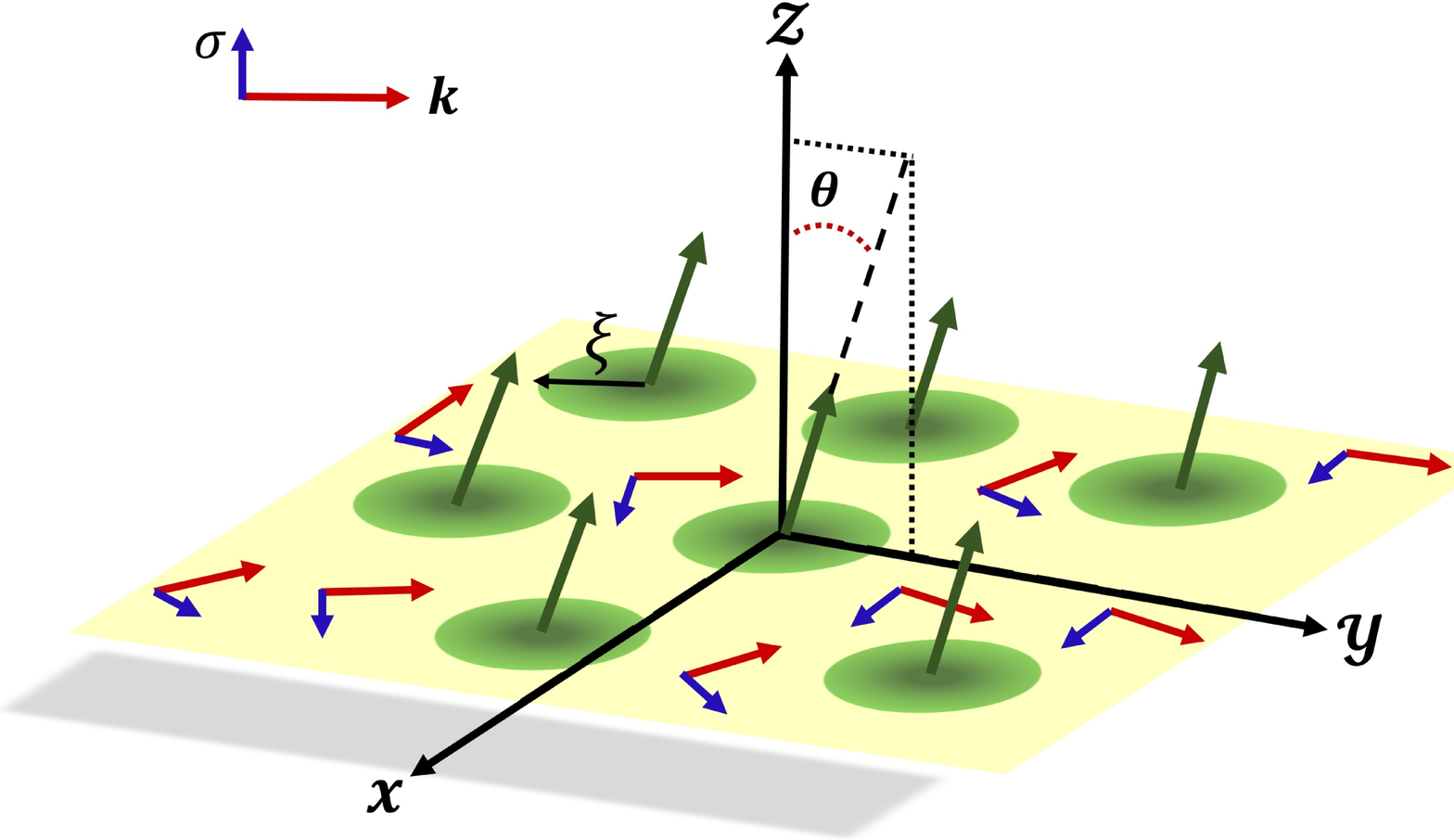}} \caption{(Color
online) The schematic illustration of the surface of a TI with
randomly distributed magnetic clusters (circles). $\xi$ is CMS,
green arrows show the spins of clusters, and $\theta$ is their tilt
angle with $z$ axis. The red and blue arrows show the momentum and
spin of Dirac electrons, respectively.}
\label{fig:Impurity-spin-direction}
\end{figure}

In TIs, doped with magnetic impurities, the interaction of
impurities with Dirac electrons affects the transport properties of
the TI. In dilute magnetic TI, the exchange interactions between
magnetic impurities (MI-MI interactions) are negligible and we can
ignore their effects on the surface conductivity of the TI. By
increasing doping, the MI-MI exchange interactions become
considerable and we should take them into account in the spin
dependent transport of the TI. The presence of MI-MI exchange
interactions, gives rise to the formation of magnetic clusters with
different sizes in the entire system. There are two rules for
constructing magnetic clusters in a spin
system\cite{KimChristensen}, one is due to purely geometric effects
which makes cluster's spins to be in the same direction and the
other is due to the correlation between the cluster's spins.
Considering such a definition of cluster, Coniglio and Klein
\cite{Coniglio:1980zz} showed that by throwing bonds between
nearest-neighbor pairs of parallel spins with a probability:
$p=1-\exp(-2J_I/k_{\rm B}T)$, the spins of the system can be divided
into clusters made of parallel spins connected by bonds. Here, $J_I$
is the nearest neighbor exchange interaction between impurities and
$k_{\rm B}$ is Boltzmann constant. The number of clusters and their
mean sizes could be computed numerically by using cluster counting
algorithms, like the Hoshen-Kopelman algorithm
\cite{PhysRevB.14.3438}, one of the most popular cluster
identification algorithm. The number of clusters and their mean
sizes depend on temperature. In general, they increase by
temperature, pass through a maximum around the critical temperature
of the surface (neglecting the back-action effects of Dirac
electrons on the formation of magnetic cluster) and decrease toward
zero at high temperatures.

In order to study the effects of MI-MI exchange interactions on the
behavior of surface conductivity, we consider magnetic clusters as
scattering centers and suppose that Dirac electrons interact with
these centers rather than single impurities\cite{PhysRevB.84.024428,
PhysRevB.77.165433}. By defining the parameter $\xi$ as the mean
size of magnetic clusters, we model the interaction of a Dirac
electron located at $\rv$ with a cluster centered at $\Rv$ as:
\begin{equation}
H_{\sigma S}=J(\rv-\Rv)\sv(\rv)\cdot\Sv(\Rv),
\label{eq:scattering-Hamiltonian}
\end{equation}
where
\begin{equation}
J(\rv-\Rv)=J_0\exp(-|\rv-\Rv|/\xi).
\label{eq:exchange-J}
\end{equation}
Here, $\Sv$ is the spin of magnetic clusters and $J_0$ is coupling
constant. The scattering potential in Eq.
(\ref{eq:scattering-Hamiltonian}) is long-range and the CMS $\xi$ is
served as the range of scattering potential. Magnetic clusters are
uncorrelated and distributed randomly on the surface of the TI.
Thus, we can investigate their effects on the surface conductivity
of the TI, separately. Without loss of generality we suppose that
clusters' spins lie on $yz$ plane, i.e. $\Sv=S (0,
\sin\theta,\cos\theta)$, where $\theta$ is the tilt angle between
$\Sv$-vector and $z$ axis (see the schematic Fig.
\ref{fig:Impurity-spin-direction}).

\section{Anisotropic relaxation times}\label{sec:relaxation-time}

In order to obtain the surface conductivity of the magnetic TI we
use semiclassical Boltzmann formalism. In the presence of an applied
electric field ${\bf E}$, the non-equilibrium distribution function
$f$ depends on the electrons' $\kv$-vector and the electric field,
satisfying the following relation\cite{Ashcroft}:
\begin{equation}
\frac{\partial f}{\partial t}=e{\bf v}_\kv\cdot{\bf
E}\left(\frac{\partial
f^0}{\partial\ve}\right),\label{eq:boltzmann1}
\end{equation}
where $e<0$, ${\bf v}_\kv=v_\kv(\cos\phi,\sin\phi)$ is the velocity
of Dirac electrons and $f^0=f^0(\ve)$ is the Fermi-Dirac
distribution function. Considering elastic scattering, and using
detailed balance $\partial f/\partial t$ can also be written as:
\begin{equation}
\frac{\partial f}{\partial t}= A \int
\frac{d^2k'}{(2\pi)^{2}}w(\kv,\kv')[f(\kv,{\bf E})-f(\kv', {\bf
E})], \label{eq:boltzmann2}
\end{equation}
where $w(\kv,\kv')$ is transition rate between the two eigenstates
($\ket{\kv}$ and $\ket{\kv'}$) of the Dirac Hamiltonian $H_0$. Using
Fermi golden rule, the transition rate is written in terms of the
scattering amplitude $|T_{\kv,\kv'}|^2$ as follows:
\begin{equation}
w(\kv,\kv')=\frac{2\pi}{\hbar}|T_{\kv,\kv'}|^2\delta(\ve_k-\ve_{k'}).\label{eq:transition-rate1}
\end{equation}
Within the first Born approximation, the $T$-matrix is given by:
$T_{\kv,\kv'}\approx\langle\kv|{\cal H}_{\sigma S}|\kv'\rangle$,
where ${\cal H}_{\sigma S}$ is the scattering Hamiltonian given by
${\cal H}_{\sigma S}=\sum_{\rv,\Rv}H_{\sigma S}$. Since magnetic
clusters are uncorrelated and distributed randomly on the surface of
the TI, we can write the transition rate as
$w(\kv,\kv')=\frac{2\pi}{\hbar}n_c|T_{\kv,\kv'}|^2\delta(\ve_k-\ve_{k'})$,
where $n_c$ is CN. By defining the parameter $\tau_\kv$ as the
momentum relaxation time of electrons and using the relaxation time
approximation: $-(f-f^0)/\tau=\partial f/\partial t$, the
non-equilibrium distribution function is given by:
\begin{equation}
f=f^0+e\tau_\kv {\bf v}_\kv\cdot{\mathbf E}\left(\frac{\partial
f^0}{\partial\ve}\right).\label{eq:boltzmann3}
\end{equation}
Making use of Eqs. (\ref{eq:boltzmann1}), (\ref{eq:boltzmann2}) and
(\ref{eq:boltzmann3}) we readily obtain:
\begin{equation}
{\bf v}_\kv\cdot{\bf E}=A\int\frac{d^2k'}{(2\pi)^2}w(\kv,\kv')(\tau_\kv{\bf v}_\kv\cdot{\bf E}-\tau_{\kv'}{\bf v}_{\kv'}\cdot{\bf E}). \label{eq:relax-time}
 \end{equation}
When the system is isotropic, (in the case of $\theta=0$), the
scattering amplitude depends on the angle of $\kv$ and $\kv'$ and
hence the relaxation time depends only on the magnitude of $\kv$ and
$\kv'$. Since $k=k'$, Eq. (\ref{eq:relax-time}) is simplified and
the relaxation time is given by the standard formula
\cite{Ashcroft}: $\frac{1}{\tau_k}=A\int
\frac{d^2k'}{(2\pi)^{2}}w(\kv,\kv')[1-\cos\Delta\phi]$, where
$\Delta\phi=\phi-\phi'$, with $\phi=\arctan(k_y/k_x)$ and $\phi
'=\arctan(k_y'/k_x')$. In the case of $\theta\neq 0$, the system is
anisotropic and the scattering amplitude as well as the relaxation
time depend on both the magnitude and directions of $\kv$ and
$\kv'$, and consequently the standard scheme is not appropriate
anymore.

The transport properties of anisotropic systems have been studied
using different
approaches\cite{PhysRevB.68.165311,PhysRevLett.99.147207,PhysRevB.75.155323,PhysRevB.79.045427}.
In this paper, we employ the recipe presented in Ref.
\onlinecite{PhysRevB.79.045427}, and approximate the non-equilibrium
distribution function, to linear order in electric field, as:
\begin{equation}
f-f^0=e E v_\kv\left(\frac{\partial f^0}{\partial
\ve}\right)[a(\phi)\cos\chi+ b(\phi)\sin\chi],\label{eq:relax-time-anisotropic}
\end{equation}
where $\chi$ is the angle between electric field and $x$ axis. Here,
the two independent functions $a(\phi)$ and $b(\phi)$ have a
dimension of time. By substituting Eq.
(\ref{eq:relax-time-anisotropic}) into Eq. (\ref{eq:boltzmann2}) and
using Eq. (\ref{eq:boltzmann1}) we achieve the following two
decoupled inhomogeneous Fredholm equations:
\begin{align}
\cos\phi=\bar{w}(\phi)a(\phi)-\int d\phi' w(\phi,\phi')a(\phi'), \label{eq:cosphi}\\
\sin\phi=\bar{w}(\phi)b(\phi)-\int d\phi' w(\phi,\phi')b(\phi'), \label{eq:sinphi}
\end{align}
where $\bar{w}(\phi)=\int d\phi' w(\phi,\phi')$ and
$w(\phi,\phi')=\frac{A}{(2\pi)^2}\int k'dk'w(\kv,\kv')$. For solving
these equations and obtaining the non-equilibrium distribution
function, we look for a solution in the form of Fourier series:
\begin{align}
a(\phi)=a_0+\sum_{n=1}\left(a^c_n\cos n \phi+ a^s_n\sin n \phi\right), \label{eq:Furier-a}\\
b(\phi)=b_0+\sum_{n=1}\left(b^c_n\cos n \phi+ b^s_n\sin n
\phi\right).\label{eq:Furier-b}
\end{align}
Putting the Fourier expansions of $a(\phi)$ and $b(\phi)$ into Eqs.
(\ref{eq:cosphi}) and (\ref{eq:sinphi}), we reach four sets of
linear recursive relations between the coefficients $a_n^c$,
$a_n^s$, $b_n^c$ and $b_n^s$. Depending on the function
$w(\phi,\phi')$, the solutions are exact or reasonably approximate.
\begin{figure}
\centerline{\includegraphics[width=80mm]{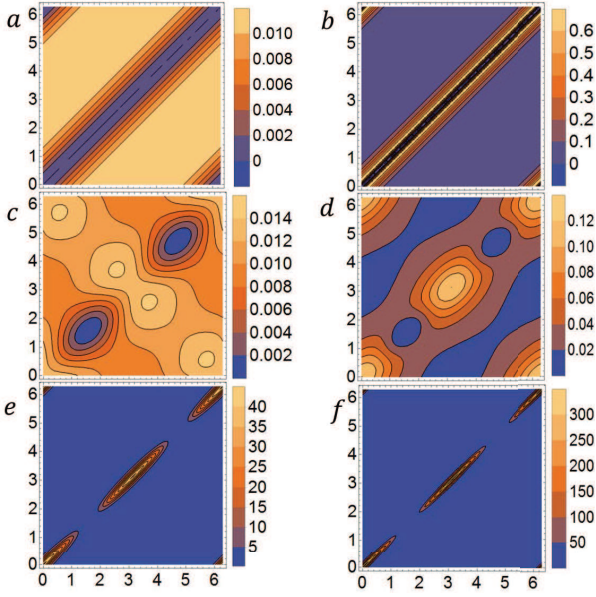}} \caption{(Color
online) Contour plots of the scattering amplitude,
$|\T_{\phi,\phi'}|^2$ versus $\phi$ (horizontal axes) and $\phi'$
(vertical axes), for different values of $k_\f\xi$ and $\theta$. In
$a$ and $b$, $k_\f\xi$ are, respectively, $0.4$ and 3, and the spins
of magnetic clusters are normal to the surface of the TI
($\theta=0$). Independent to the CMS, the forward scattering
amplitudes $|\T_{\phi,\phi}|^2$ are always zero, however the
backscattering amplitudes $|\T_{\phi,\pi+\phi}|^2$ strongly depend
on the CMS. In $c$, $d$, $f$ and $e$, the values of $k_\f\xi$ are,
respectively, $0.4$, $0.7$, 3 and 5, and the tilt angle is
$\theta=\pi/6$.  In the presence of clusters with small sizes, the
scattering amplitude is zero in the two areas around $(\phi,
\phi')=(\pi/2, \pi/2)$ and $(3\pi/2, 3\pi/2)$ (the blue areas),
which means that small clusters scatter electrons over a wide area.
When clusters become larger the areas with zero scattering amplitude
grow rapidly and electrons scatter by clusters in forward
directions.} \label{fig:T-xi-contour}
\end{figure}

Due to the elastic scattering of Dirac electrons from magnetic
clusters, the inter-band transitions are forbidden and intra-band
transitions are only allowed. Within the first Born approximation
the scattering amplitude is written as;
\begin{equation}
\begin{aligned}
&|T_{\kv,\kv'}^{++}|^{2}=|T_{\kv,\kv'}^{--}|^{2}=\frac{ 2\pi^2 J^2_0
S^2}{A^{2}}|\T_{\phi,\phi'}|^2,\\
&|\T_{\phi,\phi'}|^2=\frac{\xi^4}{(1+q^2\xi^2)^3}(1-\cos 2\theta
\cos \phi \cos \phi'-\sin \phi \sin \phi'),
\end{aligned}
\label{eq:Tkk-1}
\end{equation}
where $q=|\kv-\kv'|=k\sqrt{2(1-\cos\Delta\phi)}$. At $\theta=0$, the
spins of magnetic clusters are normal to the surface of TI and the
scattering amplitude depends only on $\Delta\phi$. At $\theta\neq
0$, the situation is however different and the scattering amplitude
depends on $\phi$ and $\phi'$ (instead of $\Delta\phi$), as shown in
Fig. \ref{fig:T-xi-contour}. In this case the system is anisotropic
which is a simultaneous effect of the spin-orbit locking of Dirac
electrons and the spin tilting of magnetic clusters. The former is
not explicitly seen in the scattering amplitude but the latter is
appeared in $\theta$.

Making use of Eqs. (\ref{eq:transition-rate1}) and (\ref{eq:Tkk-1}),
we obtain:
\begin{equation}
\begin{aligned}
w(\phi,\phi')&=\frac{1-\cos 2\theta \cos \phi \cos \phi'-\sin \phi \sin \phi'}{(1-\Omega \cos\Delta\phi)^3W_0},\\
\bar{w}(\phi)&=\frac{2+\Omega^2-3\Omega\cos^2\theta + 3\Omega\sin^2\theta\cos 2\phi}{(1-\Omega^2)^5W_0},
\label{eq:transition-rate-anisotropic}
\end{aligned}
\end{equation}
with
\begin{equation}
\Omega=\frac{2k^2\xi^2}{1+2k^2\xi^2},~~W_0=\frac{\hbar^2 v_{\rm
F}}{\pi J_0^2S^2}\frac{A}{n_c}\frac{
(1+2k^2\xi^2)^3}{k\xi^4},\label{eq:Omega and W}
\end{equation}
where $\Omega$ is a dimensionless function, and $W_0$ has a
dimension of time.

Substituting the equations (\ref{eq:Furier-a}) and
(\ref{eq:transition-rate-anisotropic}) into Eq. (\ref{eq:cosphi})
results in the following recursive relations for $a^c_i$ with odd
$i$:
\begin{equation}
\begin{aligned}
&(L_0+K_0)a^c_1+L_1 a^c_3=W_0,\\
&L_m a^c_{2m+1}+K_{m-1}a^c_{2m-1}+L_{m-1}a^c_{2m-3}=0,\quad\quad m\geq 2
\label{eq:recursive-odd-ac}
\end{aligned}
\end{equation}
where $L_m=L_m(k, \xi, \theta)$ and $K_m=K_m(k, \xi, \theta)$ are
dimensionless functions (see the Appendix, section
\ref{sec:coefficients}). By using Kramer's rule and performing some
algebraic calculation we obtain:
\begin{equation}
a^c_1=\lambda_{0}^{+}W_{0}, \label{eq:a1}
\end{equation}
with $\lambda_{0}^{+}=(L_0+K_0-\lambda_1)^{-1}$, where $\lambda_1$
is a dimensionless function, depending on $L_m$ and $K_m$ through
the recursive relation:
\begin{equation}
\lambda_m=\frac{(L_m)^2}{K_m-\lambda_{m+1}}.
\label{eq:lambdam}
\end{equation}
Since $\lambda_{0}^{+}$ converges by increasing $m$, we can
approximate $\lambda_1$ by truncating the series at some point (see
Appendix, section \ref{sec: truncation}). Higher-order coefficients
$a_i^c$ with odd $i$ are obtained from Eqs.
(\ref{eq:recursive-odd-ac}) and (\ref{eq:a1}).

The recursive relations for $a^c_i$ with even $i$ are also:
\begin{equation}
\begin{aligned}
&L'_1 a^c_2+K'_0 a^c_0=0,\\
&(L'_0+L'_1)a^c_0+K'_1a^c_2+L'_2a^c_4=0,\\
&L'_m a^c_{2m}+K'_{m-1}a^c_{2m-2}+L'_{m-1}a^c_{2m-4}=0, \quad\quad m\geq 3
\label{eq:recursive-even-ac}
\end{aligned}
\end{equation}
where the functions $L'_m=L'_m(k,\xi,\theta)$ and
$K'_m=K'_m(k,\xi,\theta)$ are presented in the Appendix, section
\ref{sec:coefficients}. The number of Dirac electrons is conserved
in scattering processes, so it is necessary the coefficient $a^c_0$
to be zero which results in $a^c_m=0$ with even $m$.

The recursive relations for odd and even $a_i^s$ are also given by:
\begin{equation}
\begin{aligned}
&(K_0-L_0)a^s_1+L_1a^s_3=0,\\
&L_ma^s_{2m+1}+K_{m-1}a^s_{2m-1}+L_{m-1}a^s_{2m-3}=0,\quad\quad m\geq 2\\
&K'_1a^s_2+L'_2a^s_4=0,\\
&L'_ma^s_{2m}+K'_{m-1}a^s_{2m-2}+L'_{m-1}a^s_{2m-4}=0.~~~m\geq 3
\label{eq:recursive-odd-even-as}
\end{aligned}
\end{equation}
One can show that all the coefficients $a_i^s$ are zero (see the
appendix). Finally, the function $a(\phi)$ is obtained in terms of
$a_i^c=a_i^c(k,\xi,\theta)$ with odd $i$ as:
\begin{equation}
a(\phi)=\sum_{n=0}^\infty a^c_{2n+1}\cos(2n+1)\phi.
\label{eq:a-phi}
\end{equation}
By analogy with the above derivation, we have also obtained all the
coefficients in $b(\phi)$  and found that except of $b^s_i$ with odd
$i$, the others are zero (see the Appendix). All these nonzero
coefficients are given in terms of $b^s_1=\lambda_0^-W_0$ with
$\lambda_0^-=(K_0-L_0-\lambda_1)^{-1}$. The function $b(\phi)$ is
thus written as:
\begin{equation}
 b(\phi)=\sum_{n=0}^{\infty} b^s_{2n+1}\sin (2n+1)\phi.
\label{eq:b-phi}
\end{equation}
Substituting the obtained functions $a(\phi)$ and $b(\phi)$ into Eq.
(\ref{eq:boltzmann3}), one can obtain the non-equilibrium
distribution function $f$.

\section{surface conductivities}\label{sec:conductivity}

In the presence of an applied electric field along $j$-direction
($E_j$), the surface conductivity of the TI along $i$-direction can
be obtained from the following relation:
\begin{equation}
\sigma_{ij}=\frac{e}{E_j}\sum_{n}\int \frac{d^2k}{(2\pi)^2}\v^{n
i}_{\kv} f(\kv,\bf E),
\end{equation}
where $n$ is the band index, $i$ and $j$ are $x$ and $y$, and $\v^{n
i}_\kv $ is the group velocity of the $n$-th band. By considering
intra-band scattering, since ${\bf v}_\kv=v_\kv(\cos\phi,\sin\phi)$,
the conductivities in $x$ and $y$ directions reduce to
\begin{align}
\sigma_{xx}=\frac{e^2}{4\pi}\int v_\kv^2\left(\frac{\partial f^0}{\partial \ve}\right)a^c_1(k,\xi,\theta,n_c) k dk,  \label{eq:sigma-xx}\\
\sigma_{yy}=\frac{e^2}{4\pi}\int v_\kv^2\left(\frac{\partial
f^0}{\partial \ve}\right)b^s_1(k,\xi,\theta,n_c) k dk.
\label{eq:sigma-yy}
\end{align}
The above relations show that the conductivities $\sigma_{xx}$ and
$\sigma_{yy}$ depend only on the coefficients $a_1^c$ and $b_1^s$,
respectively. This means that these coefficients are acting as
momentum relaxation times of Dirac electrons and play central role
on the transport properties of the system. They are actually the
effective relaxation times along $x$ and $y$ directions which depend
on the CMS $\xi$, the CN $n_c$, the tilt angle $\theta$ and the
magnitude of the incident electrons' momentum $k$. In the rest of
the paper we use the following nomenclature $\tau^x=a_1^c$ and
$\tau^y=b_1^s$.

The surface conductivities $\sigma_{xx}$ and $\sigma_{yy}$ depend on
temperature via the CMS, the CN and the derivative of the
Fermi-Dirac distribution function $\partial f^0/\partial\ve$. Using
Monte Carlo simulation one can show that the temperature dependence
of $\xi$ and $n_c$ is limited to a temperature window around a
critical temperature $T_c$ which is two orders of magnitude smaller
than the Fermi temperature $T_{\rm F}$. For example, for the three
dimensional TI ${\rm Bi}_2{\rm Se}_3$ doped with ${\rm Fe}$ atoms,
the surface is ferromagnetically ordered up to $T_c\simeq 100$ K
which is small in comparison with the Fermi temperature $T_{\rm
F}\simeq 3500$ K \cite{PhysRevB.85.195119}. The temperatures where
$\xi$ and $n_c$ vary, are much smaller than the Fermi temperature
and we can approximate the derivative of the Fermi-Dirac
distribution function, $\partial f^0/\partial\ve$, with the Dirac
delta function $\delta(\ve-\ve_{\rm F})$. With the above
considerations the surface conductivities read:
\begin{align}
\sigma_{xx}=\frac{e^2}{2h}k_\f v_\f\tau^x_{k_\f}=\sigma_0\frac{\left(1+2(k_\f\xi)^2\right)^3}{(k_\f\xi)^4}\lambda_0^+,\label{eq:conductivity-xx-anis}\\
\sigma_{yy}=\frac{e^2}{2h}k_\f
v_\f\tau^y_{k_\f}=\sigma_0\frac{\left(1+2(k_\f\xi)^2\right)^3}{(k_\f\xi)^4}\lambda_0^-,\label{eq:conductivity-yy-anis}
\end{align}
where $\sigma_0=\frac{\hbar^2v^2_\f k^4_\f A}{2\pi
n_cJ_0^2S^2}\frac{e^2}{h}$ is a function of $n_c$ with a dimension of
conductivity. The conductivity $\sigma_0$ is typically at order of
$(10^{15}-10^{17})\tilde{n}_c^{-1}e^2/h$, where $\tilde{n}_c$ is clusters density. At temperatures where CN is very large, $\tilde{n}_c$ is at order of $10^{13} m^{-2}$ and $\sigma_0\sim (10^2-10^4)\frac{e^2}{h}$. As it is seen from Eqs.
(\ref{eq:conductivity-xx-anis}) and (\ref{eq:conductivity-yy-anis}),
surface conductivities depend on the dimensionless parameter
$k_\f\xi$. This implies that the ratio of the CMS $\xi$ to the Fermi
wavelength of Dirac electrons $\lambda_\f=2\pi/k_\f$, determines the
measure of surface conductivities. As we will see below, when
$\lambda_\f$ is comparable with $\xi$, the effective relaxation
times and consequently the surface conductivities change behavior.

\subsection{Isotropic case ($\theta=0$)}\label{sec:Iso-conductivity}

When the spins of magnetic clusters are normal to the surface of the
TI, the scattering amplitude depends only on $\Delta\phi$ and the
system is isotropic. In this case the functions $L_0$ and
$\lambda_1$ are zero and the effective relaxation times $\tau^x$ and
$\tau^y$, correspond exactly to the real relaxation time
$\tau=W_0/K_0$, which could also be resulted from the standard
formula. In this case the conductivity is obtained as:
\begin{equation}
\sigma_{xx}=\sigma_{yy}=\sigma=\frac{\sigma_0}{3\pi}\frac{\left(1+4k_\f^2\xi^2\right)^{\frac
52}}{k_\f^4\xi^4}.\label{eq:conductivity-isotropic}
\end{equation}

\begin{figure}[h]
\centerline{\includegraphics[width=70mm]{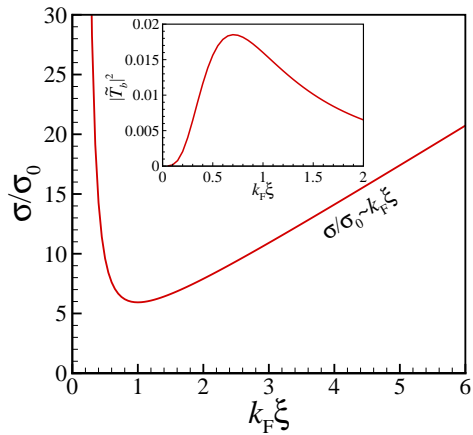}} \caption{(color online) The
surface conductivity of the TI versus $k_{\rm F}\xi$ for $\theta=0$.
Inset: backscattering amplitude at
$\theta=0$.}\label{fig:sigma-zero}
\end{figure}

While $\sigma$ has a simple dependence on $n_c$, its variation with
$\xi$ is more complicated, as shown in Fig. \ref{fig:sigma-zero}.
Note that $\sigma$ decreases initially with increasing $k_\f\xi$,
becomes minimum at a certain value of $k_\f\xi$ and then increases
at larger values of $k_\f\xi$.
Such a dependence arises mainly from the dependence of the
$T$-matrix elements, as will be explained below. The $T$-matrix is a
multiplication of two parts: one is
\begin{equation}
\T_\theta=\cos\theta[e^{i\Delta\phi}-1]+\sin\theta[e^{i\phi}-e^{-i\phi'}],\label{eq:T-theta}
\end{equation}
which depends on the direction of the clusters' spin $S$
($\theta$-dependent part) and the other is
\begin{equation}
\T_\xi=\int d \rv
e^{-i\kv\cdot\rv}e^{-r/\xi}e^{i\kv'\cdot\rv},\label{eq:T-xi}
\end{equation}
depending on the CMS ($\xi$-dependent part). For large enough values
of $\xi$ (i.e. for $\xi\gg\lambda_\f$) the scattering potential goes
to unity and $\tilde{T}_\xi$ is proportional to the Dirac delta
function $\delta(\kv-\kv')$, meaning that in the limit of infinite
$k_\f\xi$, only forward scatterings contribute to $\T_\xi$. On the
other hand, from the $\theta$-dependent part we see however that at
$\theta=0$, $\T_\theta$ is $(e^{i\Delta\phi}-1)$, which is zero for
$\Delta\phi=0$ and hence forward scattering has no contribution to
$\T_\theta$. The $T$-matrix thus goes to zero for $k_\f\xi\gg1$,
which results in an infinite surface conductivity. When CMS becomes
smaller, the conductivity decreases and becomes minimum exactly at
$k_\f\xi=1$. This dependence is mainly ascribed to the behavior of
the backscattering amplitude. Actually, when clusters shrink,
backward scattering is more likely to be occurred in the system (see
Fig. \ref{fig:T-xi-contour}-$a$ and $b$). Backscattering amplitude
increases by decreasing $\xi$ with proportion to
$|\tilde{T}_b|^2=2(k_\f\xi)^4/[1+4(k_\f\xi)^2]^3$ (see Fig.
\ref{fig:sigma-zero}, inset plot), causing reduction of
conductivity. For small values of $\xi$ (i.e. for
$\xi\ll\lambda_\f$) $\exp(-r/\xi)$ is almost zero and Dirac
electrons do not feel the scattering potential. In this limit, the
same as large-$\xi$ limit, the surface conductivity is infinite.
When clusters become larger, the scattering potential increases and
consequently the conductivity decreases first with proportion to
$(k_\f\xi)^{-4}$, passes through the minimum at $k_\f\xi=1$, and
then increases with proportion to $k_\f\xi$. This indicates that at
$\theta=0$ electrons are most efficiently scattered at particular
value of $k_\f\xi (=1)$.

The reason behind such a behavior of the surface conductivity can be
also explained as follows. The surface electron velocity operator is
directly proportional to the spin operator $\esv=\frac{\hbar}{2}\sv$
as $\v=\frac{2}{\hbar}v_{\rm F}(\hat{z}\times\esv)$. While the TI
has a vanishing equilibrium spin expectation, a finite current
density (${\bf J}=e n \la\v\ra_{neq}$) at the surface of the TI
yields a spin density $\la \esv\ra_{neq}=\frac{\hbar}{2 e v_{\rm
F}}(J_y\hat{x}-J_x\hat{y})$, where $e$ and $n$ are respectively the
electron's charge and density. According to the relation of the spin
density with current density, the surface conductivities are
proportional to the spin expectation as $\sigma_{xx}\propto\la
s_y\ra_{neq}$ and $\sigma_{yy}\propto\la s_x\ra_{neq}$.
 In the two extreme limits of $\xi\rightarrow 0$ and
$\xi\rightarrow\infty$, when the CMS is respectively very smaller
and very larger than the Fermi wavelength, actually no-scattering
happens in the system and the surface conductivity is very large.
The large conductivity for large values of $\xi$ could be explained
by the magnetization of the clusters. Actually, when magnetic
clusters are very large the magnitude of their spins is also very
large and therefore because of the exchange interaction the spin
density goes to its maximum value which results in a large surface
conductivity. When the CMS increases from zero or in the other side
decreases from a large value, the conductivity decreases and the two
branches cross at $k_\f\xi=1$. At this point since
$2\pi\xi=\lambda_{\rm F}$, the Fermi wavelength is equal to the
circumference of the cluster and an electron standing wave form
within the cluster and therefore a minimum appears on the surface
conductivity.

\subsection{Anisotropic case ($\theta\neq 0$)}\label{sec:Aniso-conductivity}

\begin{figure}[h]
\centerline{\includegraphics[width=60mm]{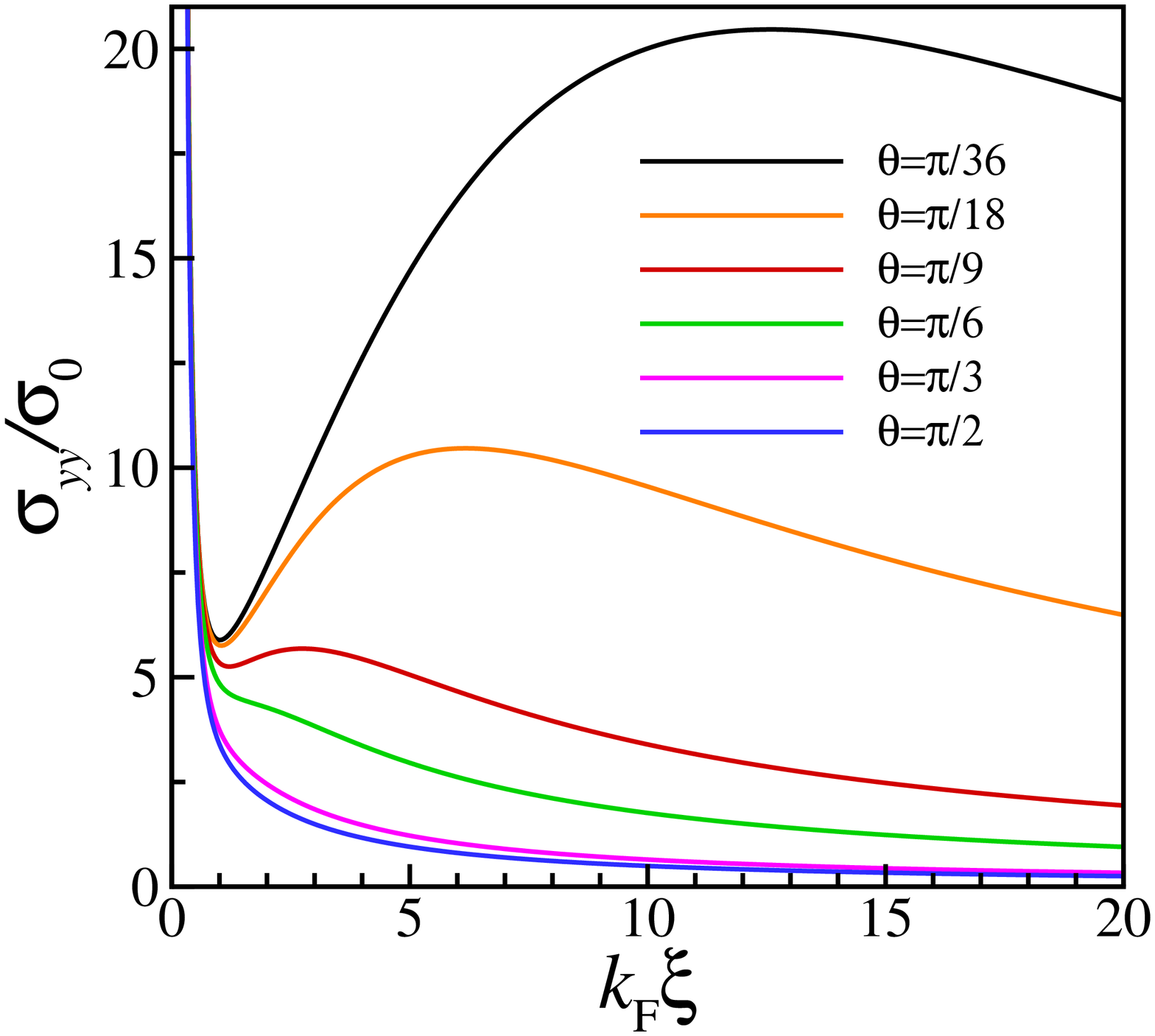}}
\centerline{\includegraphics[width=60mm]{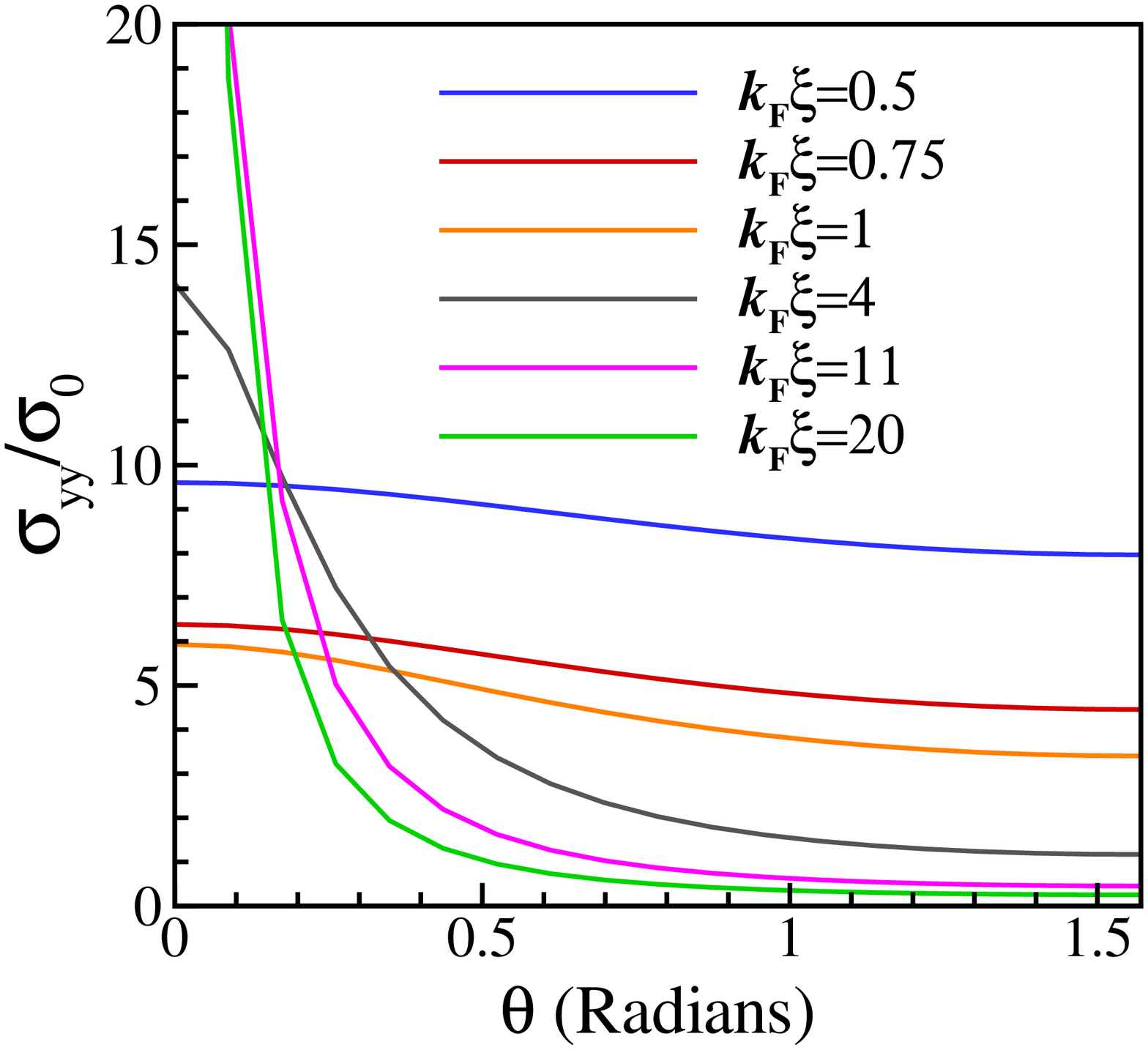}}
\caption{(Color online) The surface conductivity of the TI along $y$
direction (the direction parallel with the surface of the clusters'
spin). Top: versus $k_\f\xi$, for different values of $\theta$, and
bottom: versus $\theta$, for different CMS. For each value of
$k_\f\xi$, $\s_{yy}$ decreases monotonically by increasing $\theta$
and becomes minimum at $\theta=\pi/2$.} \label{fig:sigmayy}
\end{figure}

\begin{figure}[h]
\centerline{\includegraphics[width=60mm]{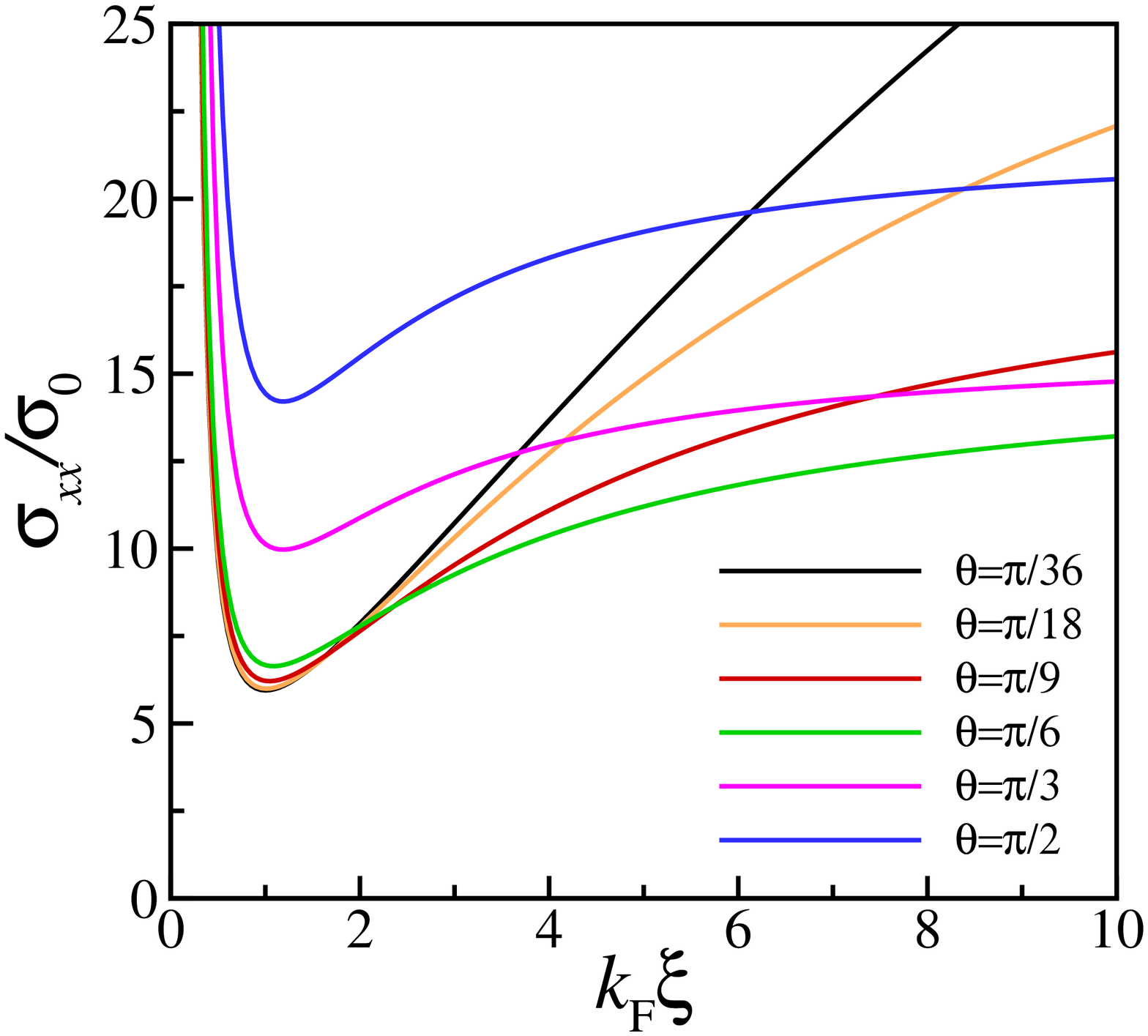}}
\centerline{\includegraphics[width=60mm]{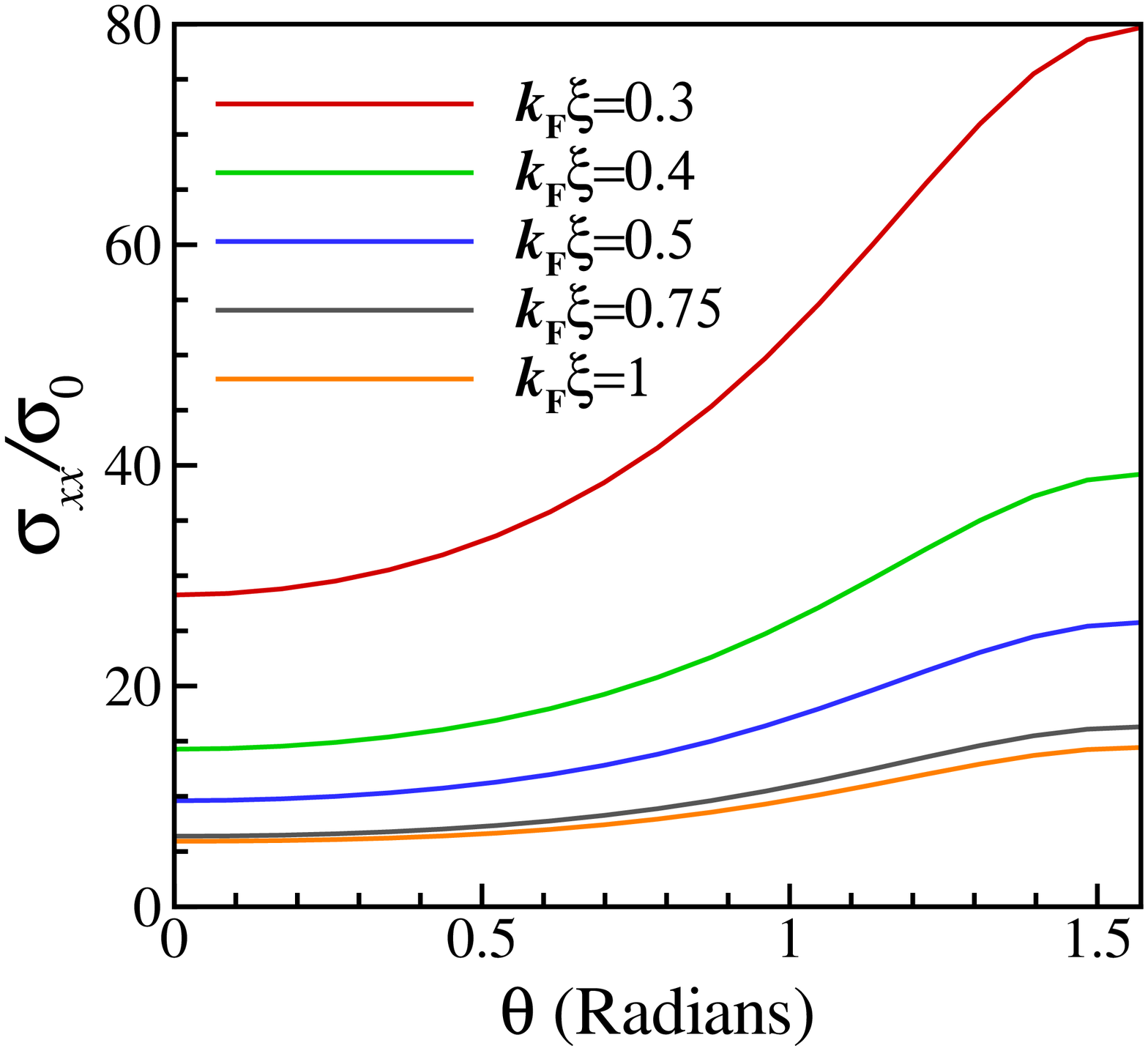}}
\centerline{\includegraphics[width=60mm]{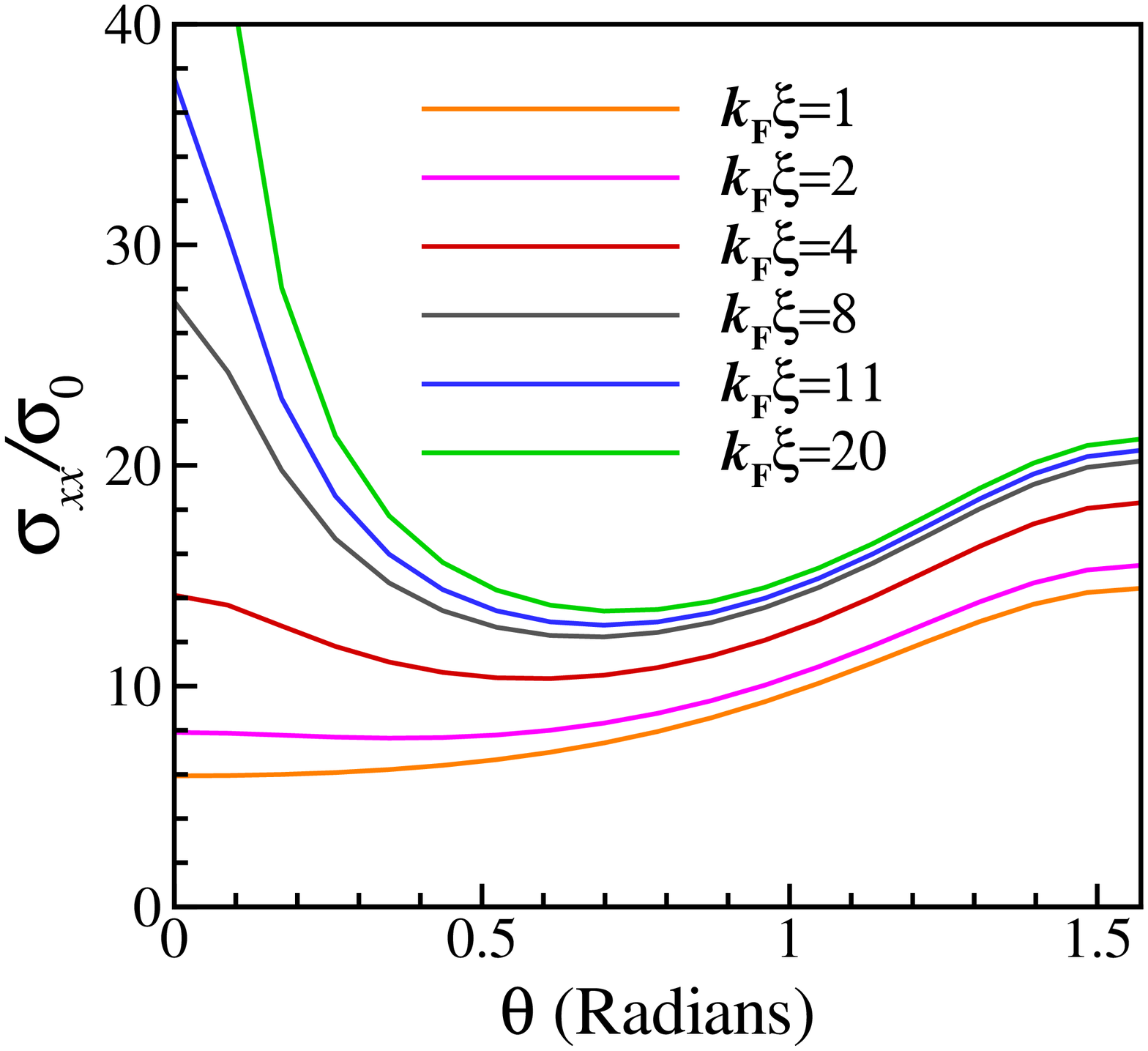}}
\caption{(Color online) The surface conductivity of the TI along $x$
direction  (the direction normal to the surface of the clusters'
spin). Top: versus $k_\f\xi$, for different values of $\theta$,
middle and bottom: versus $\theta$, for different CMS. For small
$k_\f\xi$, $\s_{xx}$ increases monotonically by increasing $\theta$
and becomes maximum at $\theta=\pi/2$. For larger $k_\f\xi$, it has
non-monotonic behavior.} \label{fig:sigmaxx}
\end{figure}

Now suppose that the spins of magnetic clusters are tilted away from
the $z$-axis along the $\theta\neq 0$ direction. In this case the
system is highly anisotropic and the surface conductivities
dependence to clusters' parameters: $\xi$ and $\theta$ is
nontrivial, as shown in Figs. \ref{fig:sigmayy} and
\ref{fig:sigmaxx}.
The same as the isotropic case, independent of the values of the
tilt angle $\theta$, the conductivity $\sigma_{xx}$ decreases
initially by increasing $\xi$, becomes minimum at $k_\f\xi=1$ and
then increases towards a finite value at $k_\f\xi\gg 1$. The
variation of $\sigma_{yy}$ by $\xi$ and $\theta$ is however
different. For $\theta\leq\pi/4$, it decreases sharply by increasing
$\xi$, becomes minimum at $k_\f\xi=1$ and then passes through a
broad maximum and finally decreases towards zero at $k_\f\xi\gg 1$.
For $\theta>\pi/4$, it decreases sharply by $\xi$, changes
dependence at $k_\f\xi=1$ and goes to zero smoothly at $k_\f\xi\gg
1$.

\begin{figure}[h]
       \centerline{\includegraphics[width=90mm]{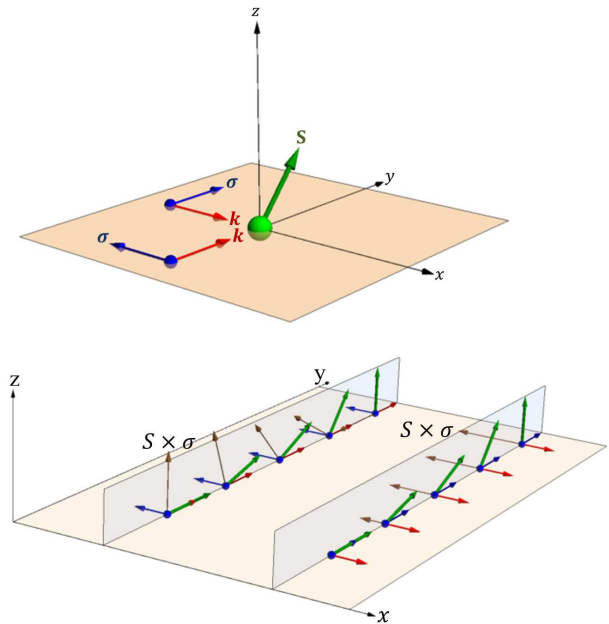}}
       \caption{(Color online) The schematic illustration of the torque exerted on the spins of Dirac electrons by a magnetic spin.
       The red and blue arrows are, respectively, the momentum and spin of Dirac electrons, the green arrows are spins of magnetic clusters and the browns are the torque exerted on electrons.
       When electrons move along $y$ direction, their spins are normal to the spin of magnetic clusters and the torque magnitude ($|\Sv\times\sv|$) does not change. However, when the electrons move along
       $y$ direction, their spins are in $y$ direction and the torque depends on the tilt angle $\theta$, decreasing by $\theta$ and becomes zero at $\theta=\pi/2$. } \label{fig:scat-torque}
\end{figure}

The non-monotonic behaviors of $\sigma_{yy}$ (the oscillatory-like
decreasing for small $\theta$, and the non-oscillatory decreasing
for larger $\theta$) are mainly ascribed to the variation of forward
and backward scattering amplitudes. For $\theta\neq 0$, the forward
and backward scattering amplitudes are simplified to
\begin{eqnarray}
&&\no|\T_f|^2=(k_\f\xi)^4(1-\cos 2\theta\cos^2\phi-\sin^2\phi),\\
&&|\T_b|^2=\frac{(k_\f\xi)^4}{\left(1+4(k_\f\xi)^2\right)^3}(1+\cos
2\theta\cos^2\phi+\sin^2\phi).
\end{eqnarray}
\begin{figure}[h]
\centerline{\includegraphics[width=130mm]{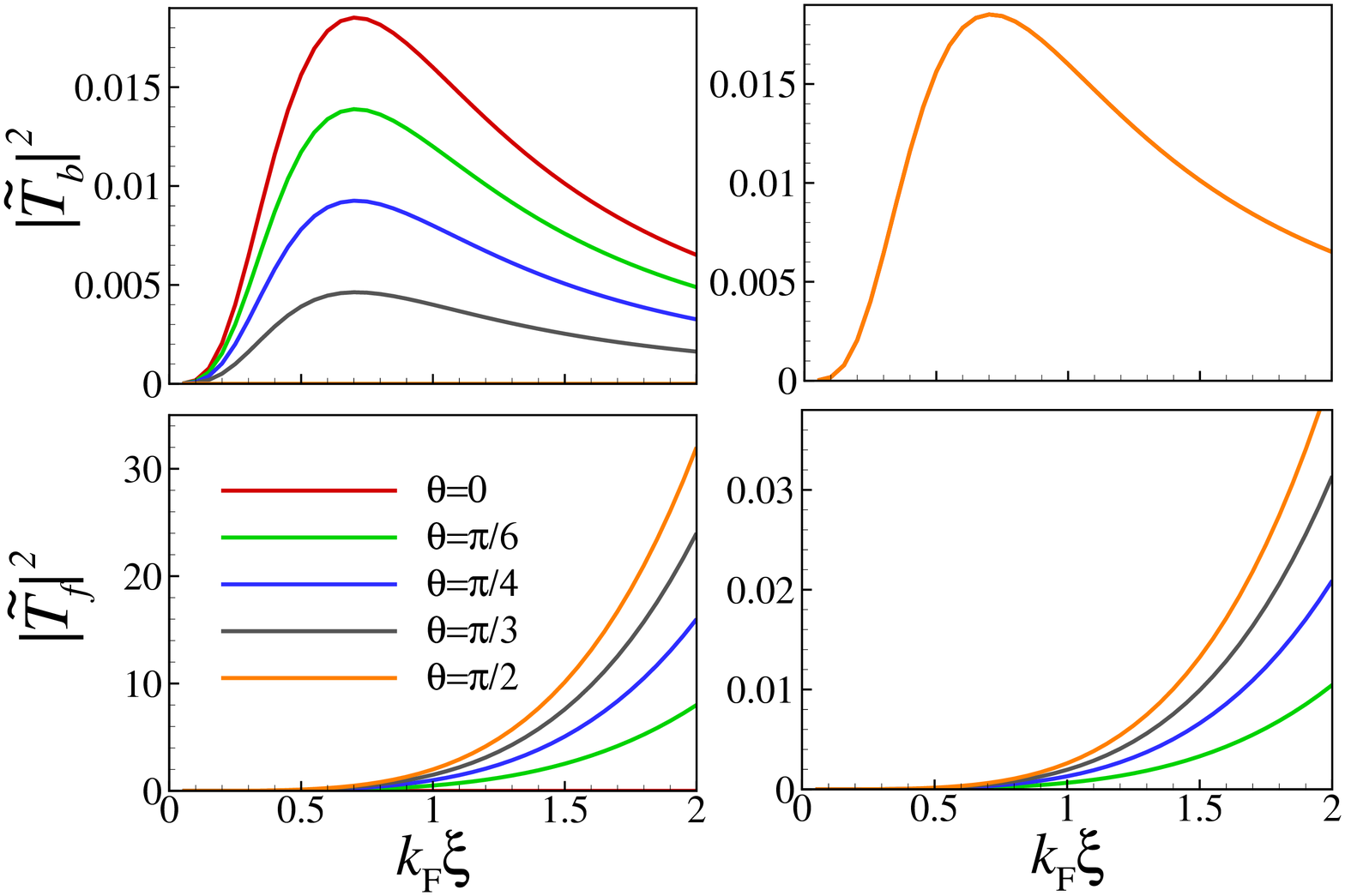}} \caption{(Color
online) The backscattering (top) and the forward scattering (bottom)
amplitudes versus $k_\f\xi$, for different values of $\theta$. Left
column: for $\phi\simeq 0$, and right column: for
$\phi\simeq\pi/2$.} \label{fig:TbTf}
\end{figure}
When the external electric field is applied along the $y$ direction,
the main contribution to the conductivity $\sigma_{yy}$ comes from
electrons whose momentums are along $y$ direction ($\phi=\pi/2$) and
hence their spins are aligned in $x$ direction. In this case, the
backscattering amplitudes are identically the same for all values of
$\theta$, as shown in Fig. \ref{fig:TbTf}, the top-right panel. It
is first zero at $\xi=0$, increases by increasing $\xi$, becomes
maximum around $k_\f\xi=1$ and then decreases gradually towards zero
by increasing $\xi$. This is in part due the fact that when the spin
of electrons lies on $x$ direction, it is normal to the $yz$ plane
and the magnetic torque exerted on the electrons $(|\Sv\times \sv|)$
does not vary by increasing the tilt angle $\theta$ (see Fig.
\ref{fig:scat-torque}).
\begin{figure}[h]
\centerline{\includegraphics[width=90mm]{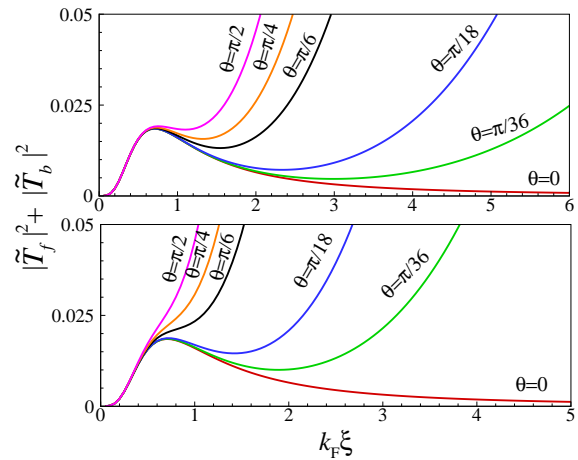}} \caption{(Color
online) The total forward and backward scattering amplitudes versus
$k_\f\xi$, for different values of $\theta$, when external electric
field is applied along $y$ direction. In this case most of
surface electrons move along $y$ direction and the incident angle is
a cute angle around $\phi=\pi/2$. Top: $\phi=92^\circ$, and bottom:
$\phi=97^\circ$.} \label{fig:TfpTb}
\end{figure}
The forward scattering amplitudes depend however on $\theta$. For
all values of $k_\f\xi$, they are zero at $\theta=0$, increase
monotonically by increasing $\theta$ and become maximum at
$\theta=\pi/2$ (see Fig. \ref{fig:TbTf}). For small values of $\xi$,
backscattered electrons have the main contribution to determine the
behavior of the conductivity $\sigma_{yy}$, whereas for larger
$\xi$, forward scattered electrons play the main role. Competition
between forward and backward scattering amplitudes specify the
dependence of the surface conductivity $\sigma_{yy}$ to $\xi$. In
order to compare these two amplitudes and to find out the origin of
the minimum/maximum appeared on $\sigma_{yy}$, we have also plotted
in Fig. \ref{fig:TfpTb}, the total amplitude $|\T_f|^2+|\T_b|^2$
versus $k_\f\xi$, for different tilt angles $\theta$. As shown in
Fig. \ref{fig:TfpTb}, the maximum at $k_\f\xi=1$ and the minimum
emerged at $k_\f\xi>1$, are respectively the reasons of the minimum
and the maximum of $\sigma_{yy}$ for $\theta>\pi/4$ and also the
reasons of changing the dependence to $k_\f\xi$ at $k_\f\xi=1$, for
$\theta\leq\pi/4$.

When the electric field is applied along the $x$ direction, the main
contribution to the transport comes from electrons whose momentums
are along $x$ axis. By increasing $\theta$ from 0 to $\pi/2$, the
magnetic torque exerted on Dirac electrons by magnetic clusters
decreases from its maximum value towards zero (see Fig.
\ref{fig:scat-torque}) and consequently the backscattering
amplitudes decrease (see Fig. \ref{fig:TbTf}, top-left panel),
causing the increase of the minimum in $\sigma_{xx}$. In contrast
with $\s_{yy}$, the conductivity $\s_{xx}$ doesn't have a monotonic
behavior with respect to $\theta$. For $k_\f\xi\leq 1$, $\s_{xx}$
increases by increasing $\theta$ and becomes maximum at
$\theta=\pi/2$ where all magnetic spins lie on the surface of the
TI. This behavior is however not seen for larger values of
$k_\f\xi$. For $k_\f\xi>1$, $\s_{xx}$ decreases first by increasing
$\theta$, passes through a minimum, roughly around $\theta=\pi/4$,
and becomes maximum at $\theta=\pi/2$. The behavior of the
conductivity $\s_{yy}$ is mainly ascribed to the behavior of the
forward and backward scattering amplitudes, however for the
conductivity $\s_{xx}$, there is no clear-cut reason and aside from
these amplitudes others have also considerable contributions in
$\s_{xx}$.

It should be noted that, while the spin of magnetic clusters exert a
torque on the spin of itinerant electrons, we expect that the
itinerant spins also conversely exert a torque on the magnetic
spins. However, in our system the spin of magnetic clusters is
approximated as a classical spin vector which has no variations in
space and time and its dynamics is much slower than that of
itinerant spins. So, the spin torque driven by the exchange
interaction only slightly modifies the damping parameter and can not
be identified as a leading mechanism for magnetic spin
damping.\cite{PhysRevLett.93.127204}


Now we compare our results with the recent theoretical studies of
the scattering of the Dirac electrons with magnetic
skyrmions.\cite{PhysRevB.96.165303, SR-Skyrmion} In Ref.
 [\onlinecite{PhysRevB.96.165303}], Araki and Nomura used a hard-wall
skyrmion approximation in which the magnetization unit vector at the
location $\rv=(\rho,\phi)$ is approximated by
$(-\sqrt{1-n_z(\rho)}\sin\phi, \sqrt{1-n_z(\rho)}\cos\phi,
n_z(\rho))$ where $n_z(\rho)=sgn(\rho-R_s)$, with $R_s$ the radius
of the skyrmion, and fully solve the scattering problem by a
hard-wall magnetic skyrmion, and estimate its effect on the
anomalous Hall conductivity using the Boltzmann transport theory.
According to this approximation the Dirac electrons move on the
surface of the TI at the presence of an out-of-plane external
magnetization along positive $\hat{z}$-direction and scattered off a
curved hard-wall. The transmitted electrons to the inside of the
skyrmion feel an out-of-plane magnetization along negative
$\hat{z}$-direction. Therefore the time reversal symmetry of the
system is broken and the band structure is gapped, so the
topological Hall effect is seen owing to the skew
scattering\cite{PhysRevB.75.045315} of the massive Dirac electron
from the hard-wall potential. Moreover,the scattering of Dirac
electrons with hard-wall skyrmions is isotropic, and therefore the
surface conductivities satisfy the relations
$\sigma_{xx}=\sigma_{yy}$ and $\sigma_{xy}=\sigma_{yx}$. In our work
magnetic clusters with tilted spins cause the system to be
anisotropic and since Dirac electrons are massless, the anomalous
Hall effect is not seen.

In Ref. [\onlinecite{SR-Skyrmion}], it has been studied numerically
the longitudinal conductance of a 3D TI/ferromagnet (FM) bilayer
where the FM supports different types of individual skyrmions.
Regarding the skyrmion as a fixed texture with its center coinciding
with the center of the FM and its spin direction given by
\begin{equation}
{\bf
m}(\rv)=(\sin\Theta(r)\cos\Phi(\phi),\sin\Theta(r)\sin\Phi(\phi),\cos\Theta(r)),\label{Eq:m}
\end{equation}
where $\Theta(r)\propto e^{-r/R_s}$, with $R_s$ being the size of
skyrmion, it has been shown that in comparison with a trivial
ferromagnetic texture where the magnetization is uniform and in the
$\hat{z}$ direction, the skyrmion textures can lead to a change of
the longitudinal resistance of the order of $k\Omega$. Moreover, the
parameter $\Delta G$ defined as the difference of the conductance of
the TI in the presence of a non-uniform skyrmion texture and a
uniform trivial texture, as a function of the input voltage $V_{in}$
depends on the FM dimensions and the skyrmion type and size. They
showed that for $V_{in}\leq 60 $ meV, $\Delta G$ linearly increases
by $\frac{\pi R_s^2}{L W}$, where $L$ and $W$ are respectively the
length and width of the FM layer. Similarly, in our system, the
surface conductivity is a nonlinear function of the CMS, however its
dependence to the CMS is more complex, originating form the
difference in the scattering potentials.

\subsection{Anisotropic magnetoresistance (AMR)}

\begin{figure}[h]
\centerline{\includegraphics[width=60mm]{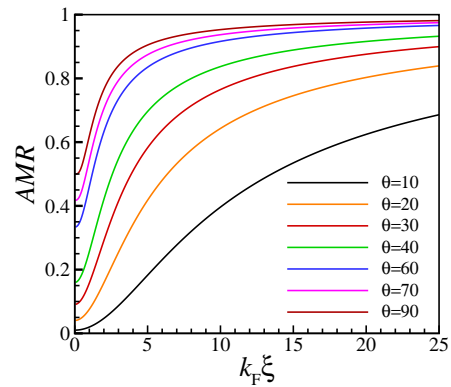}} \caption{(Color
online) Anisotropic magneto-resistance versus $k_\f\xi$, for
different values of the tilt angle $\theta$.} \label{fig:aniso}
\end{figure}

As a measure of anisotropy, we have also plotted in Fig.
\ref{fig:aniso} the AMR of the TI versus $k_\f\xi$ for different
values of the tilt angle $\theta$. Typically, AMR is a criterion for
the amount of anisotropy and is defined as \cite{PhysRevB.75.155323}
\begin{equation}
AMR=\frac{\sigma_{xx}-\sigma_{yy}}{\sigma_{xx}+\sigma_{yy}}.\label{AMR}
\end{equation}
Using Eqs. (\ref{eq:conductivity-xx-anis}) and
(\ref{eq:conductivity-yy-anis}), the AMR is obtained as:
\begin{equation}
AMR=\frac{L_0}{\lambda_1-K_0}.\label{AMR-Simple}
\end{equation}
The AMR increases gradually by increasing $k_\f\xi$ and approaches
to unity for large values of $k_\f\xi$, as shown in Fig.
\ref{fig:aniso}. At large values of $k_\f\xi$, independent of the
tilt angle $\theta$, the conductivity along $y$ direction is almost
zero and hence $AMR=1$. By increasing the tilt angle $\theta$, the
AMR also increases, indicating that the system is more anisotropic
when the clusters' spins lie on the surface of the TI. In this case,
the spins of clusters are parallel with the spin of Dirac electrons
moving along $x$-axis and is normal to the spin of electrons moving
along $y$-axis. This will make in turn the difference in the
conductivity in two directions and results in an enhancement of the
AMR by increasing $\theta$.

\begin{figure}[h]
\centerline{\includegraphics[width=60mm]{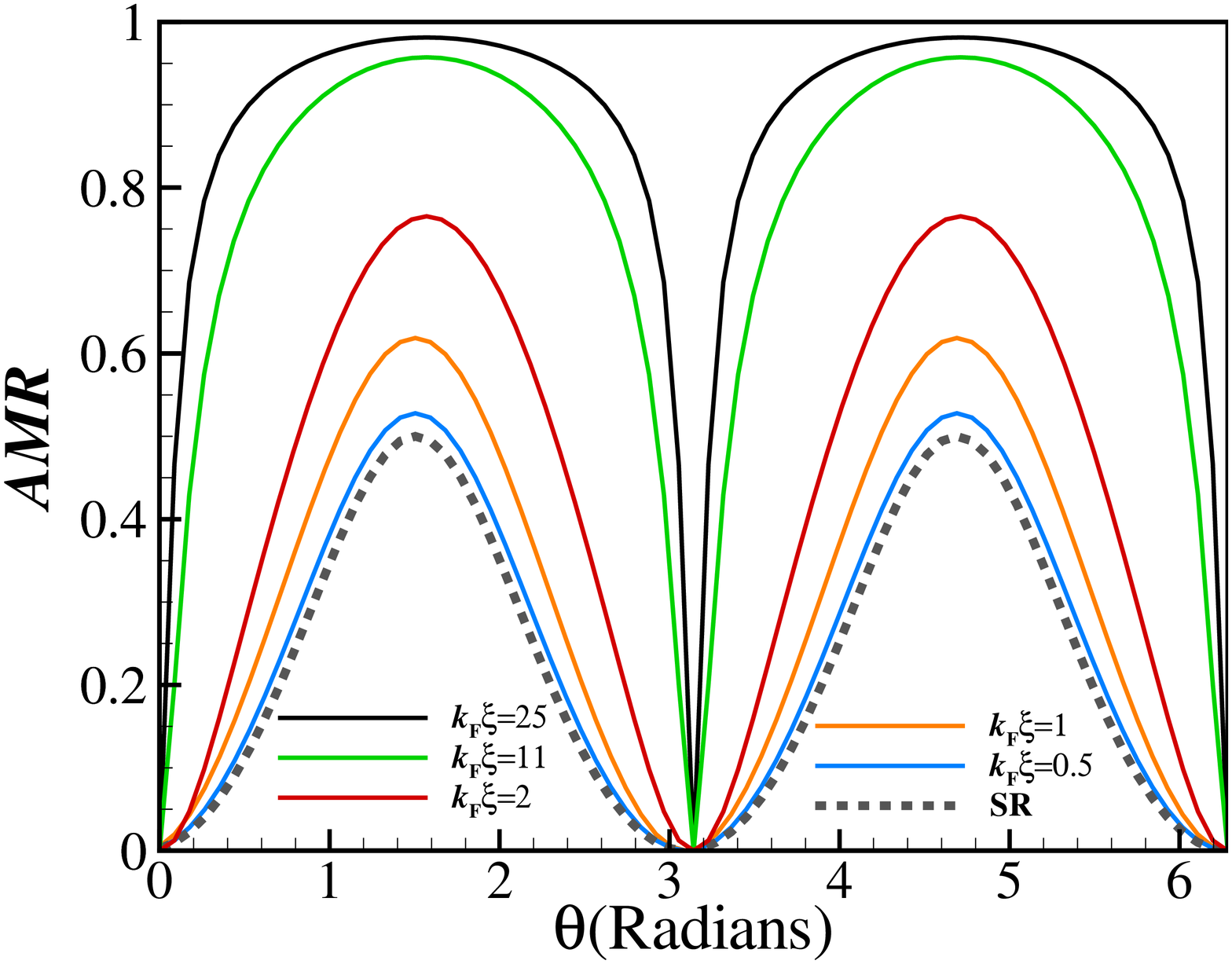}} \caption{The
angular dependence of the AMR, for different values of CMS. The bold
dotted line is the AMR of the TI doped with magnetic impurities,
where the scattering potential is the short-range Dirac delta
function.}\label{fig:AMR-theta}
\end{figure}

In Fig. \ref{fig:AMR-theta}, we have also plotted the AMR of the TI versus
the tilt angle $\theta$ for different values
of the CMS. For comparison, we have also plotted the AMR
of the magnetically doped TI where the scattering potential
is the short-range Dirac delta
function\cite{0953-8984-27-11-115301}, i.e., the interaction of the
Dirac electron located at $\rv$ with the impurity at $\Rv$ is
\begin{equation}
V=J_0\sv(\rv)\cdot\Sv(\Rv)\delta(\rv-\Rv),
\label{Eq:scat-short-range}
\end{equation}
where $J_0$ is coupling constant. In sharp contrast with our system,
in this case the surface conductivity $\sigma_{yy}$ is
$\theta$-independent, and consequently the AMR is given by
$AMR=\sin^2\theta/(2+\cos^2\theta)$, where $\theta$ is the tilt
angle of the magnetic impurities with the $z$-axis normal to the
surface of the TI.\cite{0953-8984-27-11-115301} But, as shown in
Fig. \ref{fig:AMR-theta}, taking into account the effects of
impurity-impurity interaction by means of magnetic clusters,
modifies this result significantly and the conventional angular
dependence of the AMR does not work.
In the presence of magnetic clusters, for all values of $\theta$, the AMR increases by increasing the CMS and the system experiences a giant AMR $\sim 1$ at $\theta=\pi/2$, where all magnetic spins lie on the surface of the TI.

Our results are consistent with the recent experimental work on the
AMR of the ${\rm Cr}$-doped ${\rm (Bi,Sb)_2Te_3}$ TI.\cite{AMR-Exp}
In Ref. [\onlinecite{AMR-Exp}], it has been shown that for a 1 T
magnetic field the AMR, defined as
$(\rho_{xx,max}-\rho_{xx,min})/\rho_{xx,min}$, takes a maximum value
$\sim 140\%$, and its angular dependence is completely inconsistent
with the well-known $\cos^2\theta$ angular dependence seen in
conventional ferromagnets, where $\theta$ is the tilt angle of the
magnetic field with $z$-axis. The consistency of our results with
experiment confirms that using the concept of magnetic clusters in
the TI doped with interacting magnetic impurities, properly gives
the transport properties of real magnetic TI.

\section{Nonmagnetic Impurity}\label{sec:nonmagnetic-imp}

In previous sections, we studied the surface conductivities of the
TI doped with magnetic impurities both in the isotropic ($\theta=0$)
and anisotropic ($\theta\neq 0$) cases, by defining the scattering
potential as in Eq. (\ref{eq:scattering-Hamiltonian}), composed of
two parts: an spin-dependent part ($\sv\cdot\Sv$) and a
$\xi$-dependent part ($\exp(-|\rv-\Rv|/\xi)$). We demonstrated that
the non-monotonic behavior of the surface conductivities versus
$\xi$ is a simultaneous effect of the spin-orbit locking of Dirac
electrons, and the spin direction and the CMS. In order to see the
effects of the $\xi$-dependent part on the behavior of the surface
conductivity separately, in this section we investigate the
transport properties of a 3D TI, doped with nonmagnetic impurities.
\begin{figure}[h]
\centerline{\includegraphics[width=60mm]{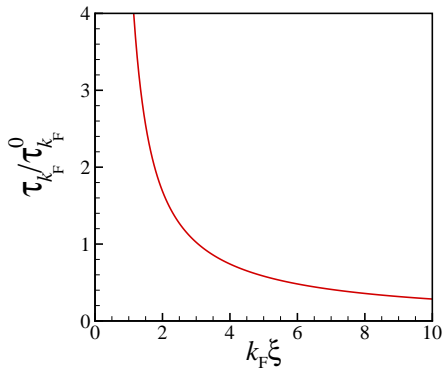}}
\caption{(Color online) The relaxation time of the TI doped with nonmagnetic impurities versus $k_\f\xi$. By increasing $\xi$, the relaxation time decreases monotonically.} \label{fig:nonmagnetic}
\end{figure}

Let us suppose that the surface of the TI is doped with non-magnetic
impurities and the scattering of an electron located at $\rv$ with
the impurity centered at $\Rv$ is given by:
\begin{equation}
V=V_0 \exp(-|\rv-\Rv|/\xi),
\end{equation}
where $V_0$ and $\xi$ are respectively the strength and the range of
the scattering potential. The scattering amplitude in this system is
given by;
\begin{equation}
|T_{\kv,\kv'}^{++}|^{2}=|T_{\kv,\kv'}^{--}|^{2}=\frac{2\pi^2
V^2_0}{A^2}\frac{\xi^4}{(1+q^2\xi^2)^3}(1+\cos\Delta\phi),
\label{eq:Tkk-non-mag}
\end{equation}
which depends only on $\Delta\phi$. The scattering is hence
isotropic and the relaxation time is given by the standard formula:
$\frac{1}{\tau_k}=A\int\frac{d^2k'}{(2\pi)^{2}}w(\kv,\kv')[1-\cos\Delta\phi]$.
Using the $T$-matrix elements in Eq. (\ref{eq:Tkk-non-mag}), the
relaxation time is obtained as:
\begin{equation}
\frac{1}{\tau_k}=\frac{1}{\tau_k^0}\frac{k^4\xi^4}{(1+2k^2\xi^2)^{3/2}},
\end{equation}
where $\tau_k^0=\frac{\hbar^2 v_\f A k^3}{\pi^2 n_c V_0^2}$ is a
parameter with a dimension of time.

We have plotted in Fig. \ref{fig:nonmagnetic}, the dimensionless
relaxation time $\tau_{k_\f}/\tau_{k_\f}^0$ of the TI doped with
nonmagnetic impurities versus $k_\f\xi$. As it is shown, it
decreases monotonically towards zero by increasing $k_\f\xi$. This
behavior is in sharp contrast with magnetic TIs. In the TIs doped
with magnetic impurities, the exchange interaction of magnetic
impurities with Dirac electrons, causes the system to be different
and the non-monotonic behaviors of conductivity (see Fig.
\ref{fig:sigma-zero}) to be seen.

\section{Summary and conclusion}\label{sec:conclusion}

We have studied the transport properties of a three dimensional
topological insulator (TI) doped with magnetic impurities by taking
the effects of impurity-impurity exchange interactions into account.
The interaction between magnetic impurities gives rise to the
formation of magnetic clusters with temperature dependent mean sizes
and numbers, randomly distributed on the surface of the TI. We have
considered that Dirac electrons scatter by magnetic clusters, rather
than single impurities, and defined the scattering potential in
terms of CMS. Making use of the semiclassical Boltzmann approach and
employing a generalized relaxation time approximation, we have
obtained the surface conductivity of the TI by solving four sets of
recursive relations. We have demonstrated that the system is highly
anisotropic and the surface conductivities strongly depend on the
mean size and the spin directions of magnetic clusters. We have
shown that, in the isotropic TI, the surface conductivity possesses
a non-monotonic behavior, it decreases by increasing CMS, passes
through a minimum where the CMS is identically the same as Fermi
wave length of Dirac electrons, and increases by increasing CMS with
proportion to the $k_\f\xi$. We have also shown that, when the
system is anisotropic, the behavior of surface conductivities with
respect to the CMS and to the clusters' spin directions is
nontrivial. Actually, independent of the values of the tilt angle of
cluster spins, the conductivity along the direction normal to the
surface of clusters' spins decreases by increasing the CMS, becomes
minimum at $k_\f\xi=1$ and then increases and saturates for larger
CMS. The behavior of the conductivity along the direction parallel
to the surface of the clusters' spins is however different. We have
shown that for tilt angles roughly smaller than $\pi/4$ it goes
oscillatory to zero by increasing CMS, and for larger tilt angles it
decreases sharply by increasing CMS, changes behavior around
$k_\f\xi=1$ and decays smoothly to zero for large CMS.

Furthermore, by comparing our results with the relaxation time and
the surface conductivity of a TI doped with long-range non-magnetic
impurities, we have demonstrated that the minimum appeared on the
surface conductivities of the magnetic TI is essentially arises from
the spin-dependent part of the potential.

We have also investigated the conductivity of two-dimensional
semiconductors in the presence of both Rashba and Dresselhous
spin-orbit couplings. Such kind of  minimum is also emerged on the
conductivity of these systems. These results will be presented in
our future works.

Since CMS and CN depend on temperature, the conductivity is also
temperature dependent. In order to investigate the behavior of the
anisotropic magneto-resistance of the magnetic TI versus
temperature, it is required to obtain the temperature dependence of
the CMS and CN. Computing these functions by means of a quantum
Monte Carlo simulation is also left for future works.

In the case that the exchange interaction $J_0$ is comparable with
the Fermi energy, the higher order terms in the T-matrix should be
taken into account on the transport properties of the system. As it
has been discussed in Ref. [\onlinecite{PhysRevB.75.045315}], the
side-jump effects which follow from the coordinate shifts during the
scattering events, and also the skew scattering which depends on the
asymmetric part of the scattering rate and is at order of
$(Scat.~Potential)^3$, have non-zero contributions to the surface
conductivities of {\it gapped} TI. In the TI with the time reversal
symmetry, these effects have vanishing contribution to the surface
conductivity and the Hall conductivity is zero. In our system, the
unperturbed Hamiltonian is gapless and skew scattering, anomalous
velocity and the side-jump effects have vanishing contributions to
the surface conductivity. However, in the case of gapped TI, where
the time reversal symmetry is broken, the system has a non-zero Hall
conductivity. Investigation of the anomalous Hall conductivity for
our system in the presence of a bulk magnetization is left for the
future study.

\begin{acknowledgments}
The authors would like to thank G. Baskaran for insightful comments
on the manuscript. Useful discussions with S. H. Abedinpour and A.
G. Moghaddam are acknowledged.
\end{acknowledgments}
\begin{widetext}

\appendix

\section{The solution of the inhomogeneous Fredholm integral equations (\ref{eq:cosphi}) and (\ref{eq:sinphi})}

In this appendix we bring the details of our calculations of the
functions $a(\phi)$ and $b(\phi)$, appeared in the non-equilibrium
distribution function $f$. For obtaining the coefficients in
$a(\phi)$ and $b(\phi)$ we should solve the integral equations
(\ref{eq:cosphi}) and (\ref{eq:sinphi}). By using Fourier series of
$a(\phi)$ and $b(\phi)$ and expand the integral kernels in terms of
trigonometric functions, the integral equations reduce to four sets
of recursive linear equations for the expansion coefficients of
Fourier series. By solving the linear equations, we obtain the
functions $a(\phi)$ and $b(\phi)$.


\subsection{Recursive relations between the coefficients appeared in $a(\phi)$ and $b(\phi)$}\label{sec:coefficients}

Employing the Fourier series of $a(\phi)$ and $b(\phi)$ (Eqs.
(\ref{eq:Furier-a}) and (\ref{eq:Furier-b})), the integrals in Eqs.
(\ref{eq:cosphi}) and (\ref{eq:sinphi}) are, respectively, written
as
\begin{eqnarray}
&&\int w(\phi,\phi')a(\phi')d\phi'=\sum_{m=0}\int w(\phi,\phi')(a_m^c\cos m\phi'+a_m^s\sin m\phi')d\phi',\label{eq:w-a}\\
&&\int w(\phi,\phi')b(\phi')d\phi'=\sum_{m=0}\int w(\phi,\phi')(b_m^c\cos m\phi'+b_m^s\sin m\phi')d\phi'.\label{eq:w-b}
\end{eqnarray}
By expanding $w(\phi,\phi')$ (see Eq.
\ref{eq:transition-rate-anisotropic}) as
\begin{equation}
w(\phi,\phi')=
 \frac{(1-\cos 2\theta \cos \phi \cos \phi'-\sin \phi \sin \phi')}{W_0}\sum_{n=0}\frac{(n+2)!}{2n!}\Omega^n\cos^n\Delta\phi,
\end{equation}
the first integral on the right hand side of Eq. (\ref{eq:w-a}) is given by:
\begin{equation}
\begin{aligned}
&\sum_{m=0}\int  w(\phi,\phi')a_m^c\cos m\phi' d\phi'\\
&=\frac{1}{W_0}\sum_{m,n=0}\int d\phi'(1-\cos 2\theta \cos \phi \cos
\phi'-\sin \phi \sin \phi')
\left(\frac{(2n+2)!}{2(2n)!}\Omega^{2n}\cos^{2n}\Delta\phi+\frac{(2n+3)!}{2(2n+1)!}\Omega^{2n+1}\cos^{2n+1}\Delta\phi\right)\times\\
&\left(a_{2m}^c\cos 2m\phi'+a_{2m+1}^c\cos(2m+1)\phi'\right)\\
&=\frac{1}{W_0}\sum_{m,n=0}\int d\phi'\bigg(\frac{(2n+2)!}{2(2n)!}\Omega^{2n}\cos^{2n}\Delta\phi\cos 2m\phi' a_{2m}^c+\frac{(2n+3)!}{2(2n+1)!}\Omega^{2n+1}\cos^{2n+1}\Delta\phi\cos(2m+1)\phi' a_{2m+1}^c\\
&-\cos 2\theta\cos\phi\Big[\frac{(2n+2)!\Omega^{2n}}{4(2n)!}\cos^{2n}\Delta\phi\big(\cos(2m+2)\phi'+\cos 2m\phi'\big)a_{2m+1}^c\\
&+\frac{(2n+3)!\Omega^{2n+1}}{4(2n+1)!}\cos^{2n+1}\Delta\phi\big(\cos(2m+1)\phi'+\cos(2m-1)\phi'\big)a_{2m}^c\Big]\\
&-\sin\phi\Big[\frac{(2n+2)!\Omega^{2n}}{4(2n)!}\cos^{2n}\Delta\phi\left(\sin(2m+1)\phi'-\sin 2m\phi'\right)a_{2m+1}^c\\
&+\frac{(2n+3)!\Omega^{2n+1}}{4(2n+1)!}\cos^{2n+1}\Delta\phi\left(\sin(2m+1)\phi'-\sin(2m-1)\phi'\right)a_{2m}^c\Big]
\bigg).
\label{eq:integral}
\end{aligned}
\end{equation}
Using the following expansions,
\[
\begin{aligned}
&\cos^{2n}\phi=\frac{1}{2^{2n}}\binom{2n}{n}+\frac{1}{2^{2n-1}}\Big[\cos 2n\phi+  \binom{2n}{1}\cos(2n-2)\phi+\dots+  \binom{2n}{n-1}\cos 2\phi\Big],\\[8pt]
& \cos^{2n-1}\phi=\frac{1}{2^{2n-1}}\Big[ \cos(2n-1)\phi+\binom{2n-1}{1}\cos(2n-3)\phi+\dots+\binom{2n-1}{n-1}\cos\phi\Big],
\end{aligned}
\]
according to the orthogonality of trigonometric functions we have \begin{equation}
\begin{aligned}
 & \int_{0}^{2\pi}\cos^{2n}\Delta\phi{\cos 2m\phi^{'} \brace \sin 2m\phi^{'}}d\phi^{'} =
  \begin{cases}
    \frac{\pi}{2^{2n-1}}\binom{2n}{n-m}{\cos 2m\phi \brace \sin 2m\phi}, & \text{ $m \leqslant n$ } \\
     0, & \text{$m\neq n$}
  \end{cases}\\[10pt]
  &\int_{0}^{2\pi}\cos^{2n+1}\Delta\phi\cos(2m+1)\phi^{'}d\phi^{'} =
    \begin{cases}
      \frac{\pi}{2^{2n}}\binom{2n+1}{n-m}{\cos(2m+1)\phi \brace \sin(2m+1)\phi}, & \text{ $m \leqslant n$ } \\
       0. & \text{$m\neq n$}
    \end{cases}
    \label{eq:m_n_integrals}
  \end{aligned}
\end{equation}
By making use of the relations in (\ref{eq:m_n_integrals}), we
obtain the integral (\ref{eq:integral}) as:
\begin{equation}
\begin{aligned}
& \sum_{m=0}\int w(\phi,\phi^{'})a_{m}^{c}\cos m\phi' d\phi'\\
&=\frac{1}{W_0}\sum_{m=0}\Big(\cos(2m-2)\phi\Big[\frac{(1-\cos 2\theta)}{4}G(m-1,\Omega)\Big]a_{2m}^{c}
+\cos(2m-1)\phi \Big[F(m,\Omega)(1-\cos 2\theta)\Big]a_{2m+1}^{c}\\
&+ \cos 2m\phi\Big[ 4F(m,\Omega)-\frac{(1+\cos 2\theta)}{4}(G(m-1,\Omega)+G(m,\Omega))\Big]a_{2m}^{c}\\
&+ \cos(2m+1)\phi\Big[G(m+1,\Omega)-(1+\cos 2\theta)(F(m,\Omega)+F(m+1,\Omega))\Big]a_{2m+1}^{c}\\
&+ \cos(2m+2)\phi\Big[\frac{(1-\cos
2\theta)}{4}G(m,\Omega)\Big]a_{2m}^{c} +\cos(2m+3)\phi\Big[(1-\cos
2\theta)F(m+1,\Omega)\Big]a_{2m+1}^{c}\Big),
  \end{aligned}
  \label{eq:cosine_a_integral}
\end{equation}
where
\begin{equation}
\begin{aligned}
F(m,\Omega)=\sum_{n=0}\frac{(2n+2)!}{4(n-m)!(n+m)!}\left(\frac{\Omega}{2}\right)^{2n}
=&\sum_{k=0}\frac{(2m+2k+2)!}{4k!(2m+k)!}\left(\frac{\Omega}{2}\right)^{2m+2k}\\
=&\frac{(1+m)(1+2m)}{2}\left(\frac{\Omega}{2}\right)^{2m}{}_2F_{1}[\frac{3}{2}+m,2+m,1+2m,\Omega^{2}],\\
G(m,\Omega)=\sum_{n=0}\frac{(2n+3)!}{(n-m)!(n+m+1)!}\left(\frac{\Omega}{2}\right)^{2n+1}
=&\sum_{k=0}\frac{(2m+2k+3)!}{k!(2m+k+1)!}\left(\frac{\Omega}{2}\right)^{2m+2k+1}\\
=&2m(1+2m)\left(\frac{\Omega}{2}\right)^{2m+1}{}_2F_{1}[2+m,\frac{5}{2}+m,2+2m,\Omega^{2}].
\end{aligned}
\label{eq:hyper}
\end{equation}
Here, ${}_2F_{1}[\frac 32+m,2+m,1+2m,\Omega^2]$ and
${}_2F_{1}[2+m,\frac 52+m,2+2m,\Omega^2]$ are hypergeometric
functions. Since, $0\leq \Omega \leq 1$, the functions $F(m,\Omega)$
and $G(m,\Omega)$ are simplified as;
\begin{equation}
\begin{aligned}
&F(m,x)=\frac{\pi(2+x^{2}+6m\sqrt{1-x^{2}}-4m^{2}(1-x^{2}))}{4 x^{-2m}(1+\sqrt{1-x^{2}})^{2m}(1-x^{2})^{\frac{5}{2}}},\\
& G(m-1,x)=\frac{3\pi-3\pi\sqrt{1-x^{2}}(1-2m)+4\pi m(m-1)(1-x^{2})}{x^{1-2m}(1+\sqrt{1-x^{2}})^{2m-1}(1-x^{2})^{\frac{5}{2}}}.
\end{aligned}
\end{equation}

The second integral on the right hand side of Eq. (\ref{eq:w-a}) is also obtained as:
\begin{equation}
\begin{aligned}
&\sum_{m=0}\int w(\phi,\phi')a_m^s\sin m\phi' d\phi'\\
&=\frac{1}{W_0}\sum_{m=0}\Big(\sin(2m-2)\phi\Big[\frac{(1-\cos 2\theta)}{4}G(m-1,\Omega)\Big]a_{2m}^{s}
+\sin(2m-1)\phi \Big[F(m,\Omega)(1-\cos 2\theta)  \Big]a_{2m+1}^{s}\\
&+\sin 2m\phi\Big[ 4F(m,\Omega)-\frac{(1+\cos 2\theta)}{4}(G(m-1,\Omega)+G(m,\Omega))\Big]a_{2m}^{s}\\
&+\sin(2m+1)\phi\Big[G(m+1,\Omega)-(1+\cos 2\theta)(F(m,\Omega)+F(m+1,\Omega))\Big]a_{2m+1}^{s}\\
&+\sin(2m+2)\phi\Big[\frac{(1-\cos
2\theta)}{4}G(m,\Omega)\Big]a_{2m}^{s} +\sin(2m+3)\phi\Big[(1-\cos
2\theta)F(m+1,\Omega)\Big]a_{2m+1}^{s}\Big).
  \end{aligned}
   \label{eq:sine_a_integral}
\end{equation}

Substituting (\ref{eq:cosine_a_integral}) and
(\ref{eq:sine_a_integral}) into Eq. (\ref{eq:cosphi}), and
equalizing the coefficients of the trigonometric functions with the
same arguments, we reach the recursive relations in Eqs.
(\ref{eq:recursive-odd-ac}), (\ref{eq:recursive-even-ac}) and
(\ref{eq:recursive-odd-even-as}) which are respectively expressed in
the following matrix forms:
\begin{equation}
    \begin{bmatrix} L_{0}+K_{0}& L_{1}& 0& 0& 0&\dots\\
   L_{1}&K_{1}& L_{2} & 0& 0 & 0&\dots\\
   0&L_{2} & K_{2} &L_{3} & 0 & 0 &\dots\\
   0&0&L_{3} & K_{3} &L_{4} &0&\dots\\
   0&0&0& L_{4} & K_{4} & L_{5} &\dots\\
   \vdots&\vdots&\vdots&\vdots&\vdots&\ddots
     \end{bmatrix}
     \begin{bmatrix}a^{c}_{1}\\
       a^{c}_{3}\\
       a^{c}_{5}\\
       a^{c}_{7}\\
       a^{c}_{9}\\
       \vdots
        \end{bmatrix}
        =
       \begin{bmatrix}W_{0}\\
            0\\
            0\\
            0\\
            0\\
            \vdots
             \end{bmatrix},
             \label{ma}
  \end{equation}
       \begin{equation}
             \begin{bmatrix} K_{0}-L_{0}&L_{1}&0&0&0&\dots\\
            L_{1}&K_{1}&L_{2}&0&0&0&\dots\\
            0&L_{2}&K_{2}&L_{3}&0&0&\dots\\
            0&0&L_{3}&K_{3}&L_{4}&0&\dots\\
            0&0&0&L_{4}&K_{4}&L_{5}&\dots\\
            \vdots&\vdots&\vdots&\vdots&\vdots&\ddots
              \end{bmatrix}
              \begin{bmatrix}a^{s}_{1}\\
                a^{s}_{3}\\
                a^{s}_{5}\\
                a^{s}_{7}\\
                a^{s}_{9}\\
                \vdots
                 \end{bmatrix}
                 =
                \begin{bmatrix}0\\
                     0\\
                     0\\
                     0\\
                     0\\
                     \vdots
                      \end{bmatrix},
                      \label{m3}
           \end{equation}
and
      \begin{equation}
        \begin{bmatrix} K_{1}^{'}&L_{2}^{'}&0&0&0&\dots\\
       L_{2}^{'}&K_{2}^{'}&L_{3}^{'}&0&0&0&\dots\\
       0&L_{3}^{'}&K_{3}^{'}&L_{4}^{'}&0&0&\dots\\
       0&0&L_{4}^{'}&K_{4}^{'}&L_{5}^{'}&0&\dots\\
       0&0&0&L_{5}^{'}&K_{5}^{'}&L_{6}^{'}&\dots\\
       \vdots&\vdots&\vdots&\vdots&\vdots&\ddots
         \end{bmatrix}
         \begin{bmatrix}a^{s}_{2}\\
           a^{s}_{4}\\
           a^{s}_{6}\\
           a^{s}_{8}\\
           a^{s}_{10}\\
           \vdots
            \end{bmatrix}
            =
           \begin{bmatrix}0\\
                0\\
                0\\
                0\\
                0\\
                \vdots
                 \end{bmatrix},
                 \label{m4}
       \end{equation}
where
\begin{equation}
\begin{aligned}
L_{m}(k,\theta,\xi)&=\frac{3\pi\Omega \sin^{2}\theta}{2(1-\Omega^{2})^{\frac{5}{2}}}+\big(\cos 2\theta-1\big) F(m,\Omega),\\
K_{m}(k,\theta,\xi)&=\frac{\pi\big(2+\Omega^{2}-3\Omega\cos^{2}\theta\big)}{(1-\Omega^{2})^{\frac{5}{2}}}-G(m+1,\Omega)
+\big(1+\cos 2\theta\big)\big(F(m,\Omega)+F(m+1,\Omega)\big),
\label{eq:Lm-Km}
\end{aligned}
\end{equation}
and
\begin{equation}
\begin{aligned}
L'_m(k,\theta,\xi)&=\frac{3\pi\Omega \sin^{2}\theta}{2(1-\Omega^{2})^{\frac{5}{2}}}+ \frac 14(\cos 2\theta-1\big)G(m,\Omega),\\
K'_m(k,\theta,\xi)&=\frac{\pi\big(2+\Omega^2-3\Omega\cos^2\theta\big)}{2(1-\Omega^2)^{\frac 52}}-4 F(m,\Omega)+\frac 14\big(1+\cos 2\theta\big)\big(G(m,\Omega)+G(m+1,\Omega)\big).
\label{eq:Lmp-Kmp}
 \end{aligned}
 \end{equation}
By analogy with the above derivations, using Eq. (\ref{eq:sinphi}),
the recursive relations between the coefficients appeared in
$b(\phi)$ are also obtained as
 \begin{equation}
 \begin{aligned}
 &(L_0+K_0)b^c_1+L_1 b^c_3=0,\\
 &L_m b^c_{2m+1}+K_{m-1}b^c_{2m-1}+L_{m-1}b^c_{2m-3}=0,\quad\quad m\geq 2
 \end{aligned}
 \label{eq:recursive-odd-bc}
 \end{equation}
 \begin{equation}
 \begin{aligned}
 &L'_1 b^c_2+K'_0 b^c_0=0,\\
 &(L'_0+L'_1)b^c_0+K'_1 b^c_2+L'_2 b^c_4=0,\\
 &L'_m b^c_{2m}+K'_{m-1} b^c_{2m-2}+L'_{m-1} b^c_{2m-4}=0, \quad\quad m\geq 3
 \end{aligned}
 \label{eq:recursive-even-bc}
 \end{equation}
\begin{equation}
 \begin{aligned}
 &(K_0-L_0) b^s_1+L_1 b^s_3=W_0,\\
 &L_m b^s_{2m+1}+K_{m-1} b^s_{2m-1}+L_{m-1} b^s_{2m-3}=0,\quad\quad m\geq 2\\
 &K'_1b^s_2+L'_2b^s_4=0,\\
 &L'_m b^s_{2m}+K'_{m-1} b^s_{2m-2}+L'_{m-1} b^s_{2m-4}=0.~~~m\geq 3
 \end{aligned}
 \label{eq:recursive-odd-even-bs}
 \end{equation}
Solving the above recursive relations, results in the functions
$a(\phi)$ and $b(\phi)$.


\subsection{Truncation}\label{sec: truncation}

In order to obtain the functions $a(\phi)$ and $b(\phi)$, we have to
truncate the series in Eqs. (\ref{eq:w-a}) and (\ref{eq:w-b}) at
some point. The accuracy of the results depends on the number of
independent trigonometric functions we kept in the series. By
keeping $l$ number of functions, the matrix equation in (\ref{ma})
reduces to:
  \begin{equation}
   \begin{bmatrix} L_{0}+K_{0}& L_{1}& 0& 0&\dots& 0\\
  L_{1}&K_{1}& L_{2} & 0& \dots & 0\\
  0&L_{2} & K_{2} &L_{3} & \dots & \vdots\\
  \vdots&\vdots&\vdots&\vdots&\vdots& 0\\
  0 & \dots &0& L_{l-1} & K_{l-1}& L_l\\
  0& \dots & 0 & 0 & L_l & K_l
    \end{bmatrix}
    \begin{bmatrix}a^{c}_{1}\\
      a^{c}_{3}\\
      a^{c}_{5}\\
      \vdots \\
      a^{c}_{2l-1}\\
      a^{c}_{2l+1}\\
      \end{bmatrix}
       =
      \begin{bmatrix}W_{0}\\
           0\\
           0\\
           \vdots\\
           0\\
           0\\
           \end{bmatrix}.
            \label{fma}
 \end{equation}
\begin{figure}
\centerline{\includegraphics[width=60mm]{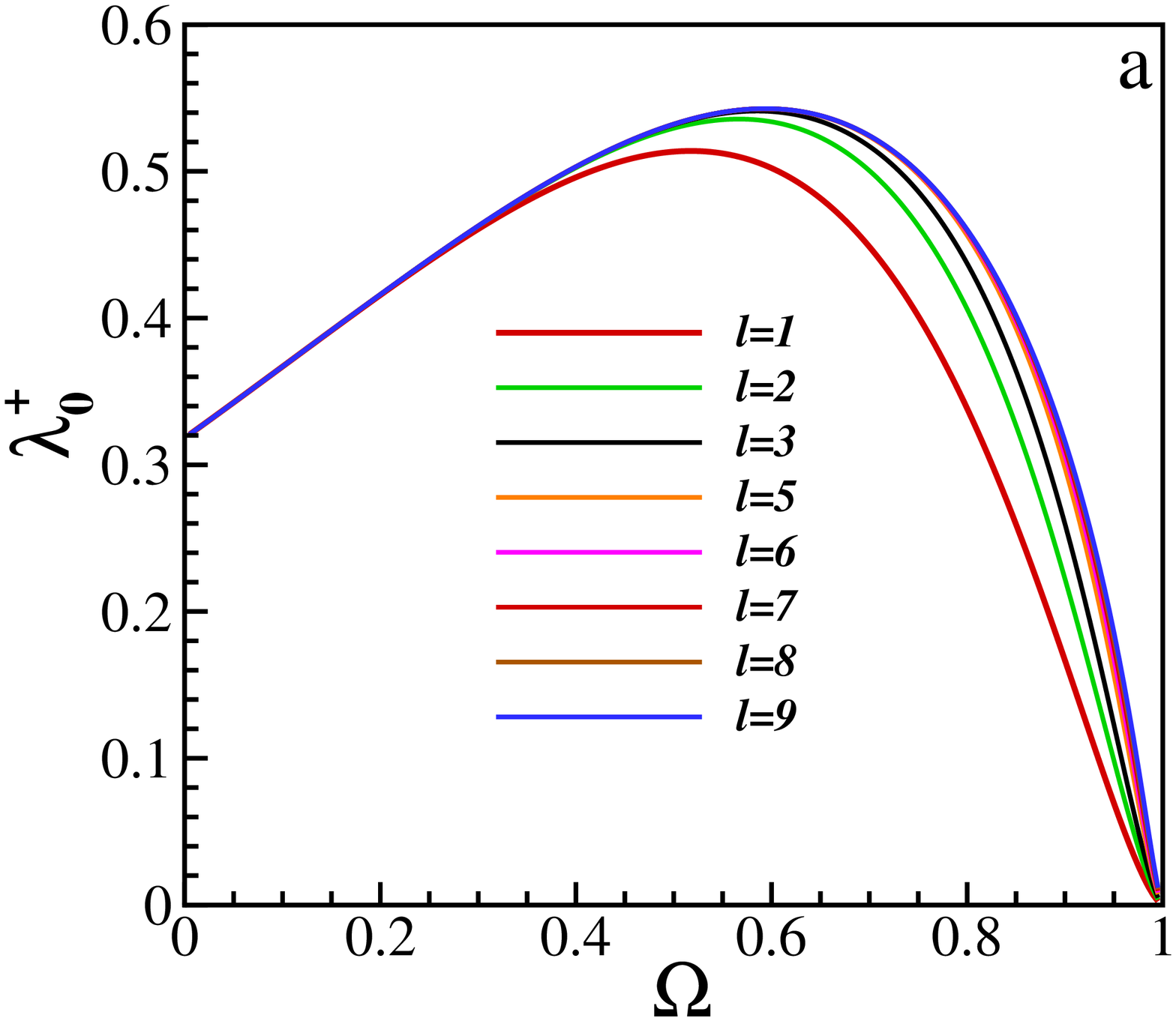}
\includegraphics[width=60mm]{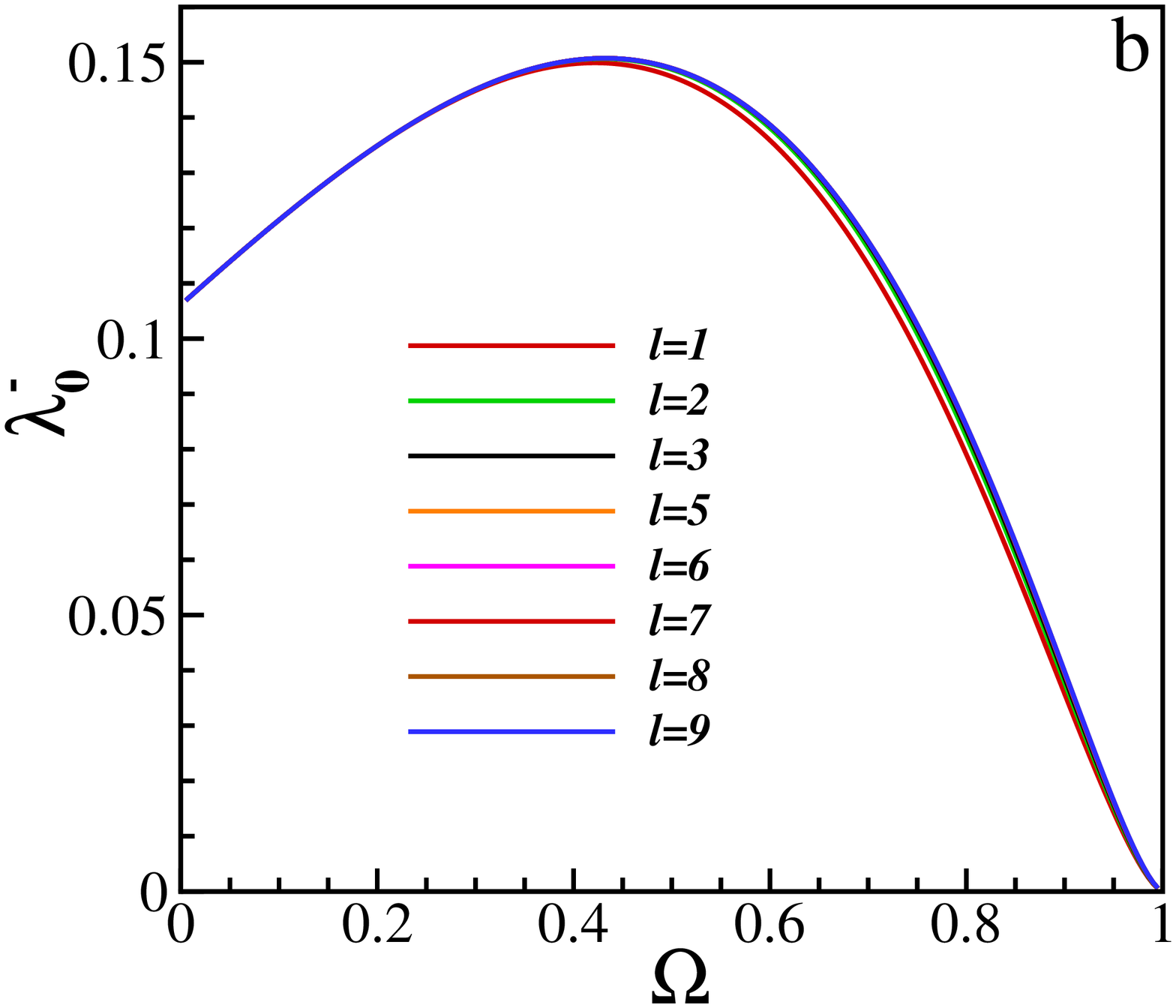}}
\caption{(Color online) The functions $\lambda_0^+$ (a) and
$\lambda_0^-$ (b) versus $\Omega$, for different values of $l$, in
the case of highly anisotropic system ($\theta=\pi/2$).}
\label{fig:denominator}
\end{figure}
The coefficient $a_1^c$ can be obtained from the relation
$a_1^c=\det D_1/\det D$, where $D$ is the square matrix on the left
hand side of Eq. (\ref{fma}) and $D_1$ is a square matrix, obtaining
from the $D$-matrix by replacing the first column with the column
vector on the right hand side of the Eq. (\ref{fma}). Computing the
determinant of $D_1$ and $D$, the coefficient $a_1^c$ reads:
\begin{equation}
a_{1}^{c}=\frac{W_{0}}{(L_{0}+K_{0})-L_{1}^{2}\frac{d_{2}}{d_{1}}}=\frac{W_{0}}{(L_{0}+K_{0})-\frac{L_{1}^{2}}{K_{1}-L_{2}^{2}\frac{d_{3}}{d_{2}}}}=\\
\frac{W_{0}}{(L_{0}+K_{0})-\frac{L_{1}^{2}}{K_{1}-\frac{L_{2}^{2}}{K_{2}-L_{3}^{2}\frac{d_{4}}{d_{3}}}}}=\dots,
\label{determinant}
\end{equation}
where $d_n$ is the following determinant:
   \begin{equation}
   \begin{aligned}
   d_{n}&= \begin{vmatrix}
   K_{n}&L_{n+1}&0&0&0&\dots&0\\
   L_{n+1}&K_{n+1}&L_{n+2}&0&0&\dots&0\\
   0&L_{n+2}&K_{n+2}&L_{n+3}&0&\dots&0\\
   0&0&\vdots&\vdots&\vdots&\dots&0\\
   \vdots&\vdots&\vdots&\vdots&\vdots&\vdots&0\\
   \vdots&\vdots&\vdots& L_{l-1}&K_{l-1}& L_l& 0\\
   0 & 0& 0&\dots& 0& L_l& K_l
   \end{vmatrix},
   \end{aligned}
   \end{equation}
and $\frac{d_{n}}{d_{n-1}}=(K_{n-1}-L_{n}^{2}\frac{d_{n+1}}{d_{n}})^{-1}$.
By defining
\[ \lambda_{1}=\cfrac{L_{1}^{2}}{K_{1} -\cfrac{L_{2}^{2}}{K_{2} -\cfrac{L_{3}^{2}}{
      \begin{array}{@{}c@{}c@{}c@{}}
        K_{3} - {}\\ &\ddots\\ &&{} K_{l-1}- \cfrac{L_l^2}{K_l}
      \end{array}
    }}}~ ,\]
the coefficient $a_1^c$ is written in the following compact form:
 \begin{equation}
    \begin{aligned}
           a^{c}_{1}=\frac{W_{0}}{L_{0}+K_{0}-\lambda_{1}}=\tilde{\tau}_k\frac{(1+2k^2\xi^2)^3}{k^4\xi^4}\lambda_{0}^{+}.
     \label{a1ct}
       \end{aligned}
    \end{equation}
where $\tilde{\tau}_k=\frac{\hbar^2 v_\f A k^3}{\pi n_c J_0^2S^2}$
is a constant with a dimension of time. We have plotted in Fig.
\ref{fig:denominator}-a, the parameter
$\lambda_0^+=(K_{0}+L_{0}-\lambda_{1})^{-1}$, versus $\Omega$, for
different values of $l$. As it is shown, for $\Omega<0.5$ (i.e. for
$k_\f\xi<1$) the function $\lambda_0^+$ is independent of $l$ and a
precise closed form expression for $\lambda_0^+$ is obtained by
keeping only $l=2$ number of independent trigonometric functions in
the series. For $\Omega>0.5$ or $k_\f\xi>1$, the situation is
however different and the number $l$ is crucial to reach to the
precise value of $\lambda_0^+$. For an arbitrary $\Omega (>0.5)$,
the parameter $\lambda_0^+$ increases by increasing $l$ and
saturates rapidly at a finite $l$. Thus by employing a finite number
of independent functions (for example $l\sim9$), we obtain a closed
form expression for $\lambda_0^+$.
\begin{figure}
\centerline{\includegraphics[width=60mm]{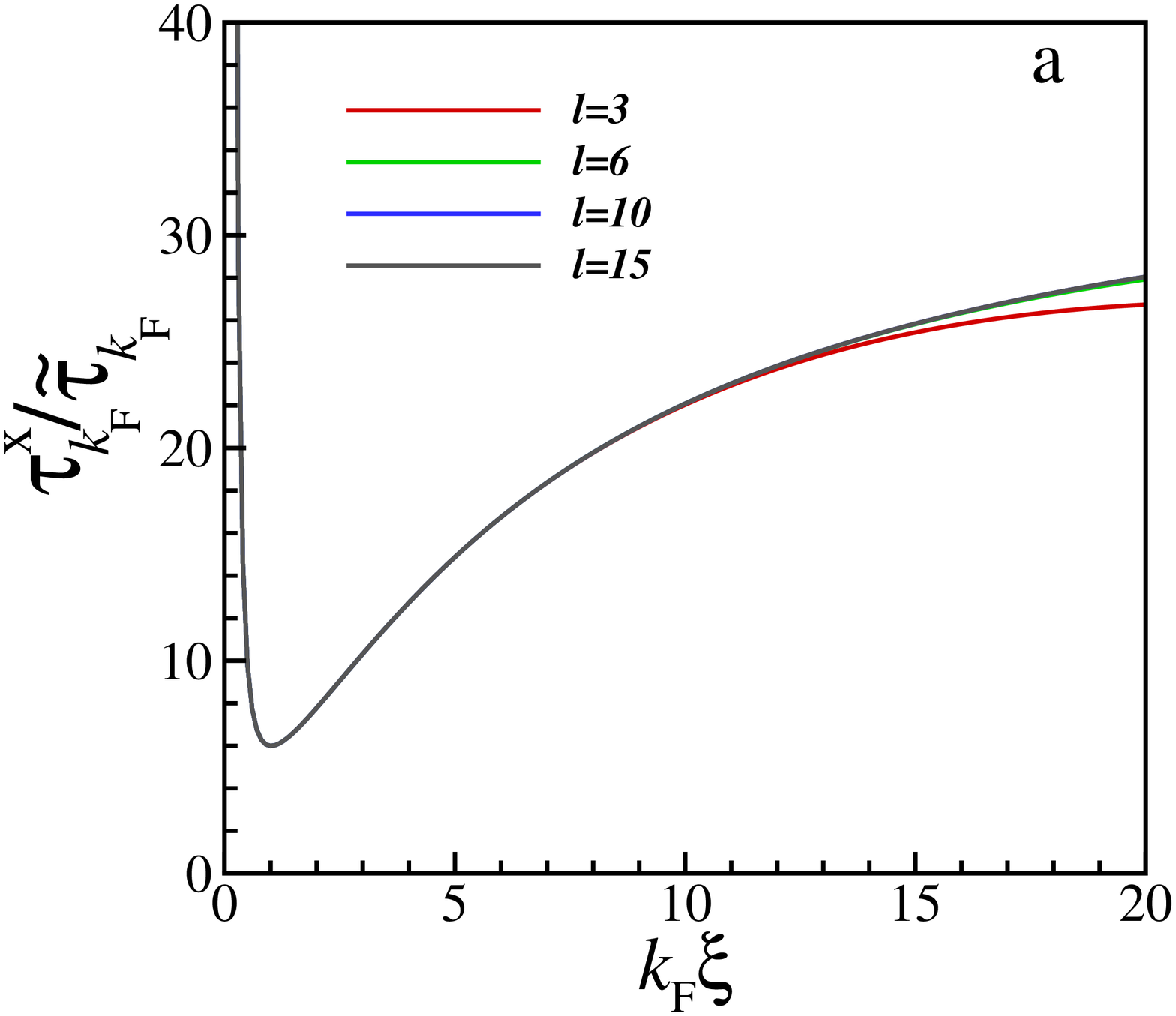}
\includegraphics[width=60mm]{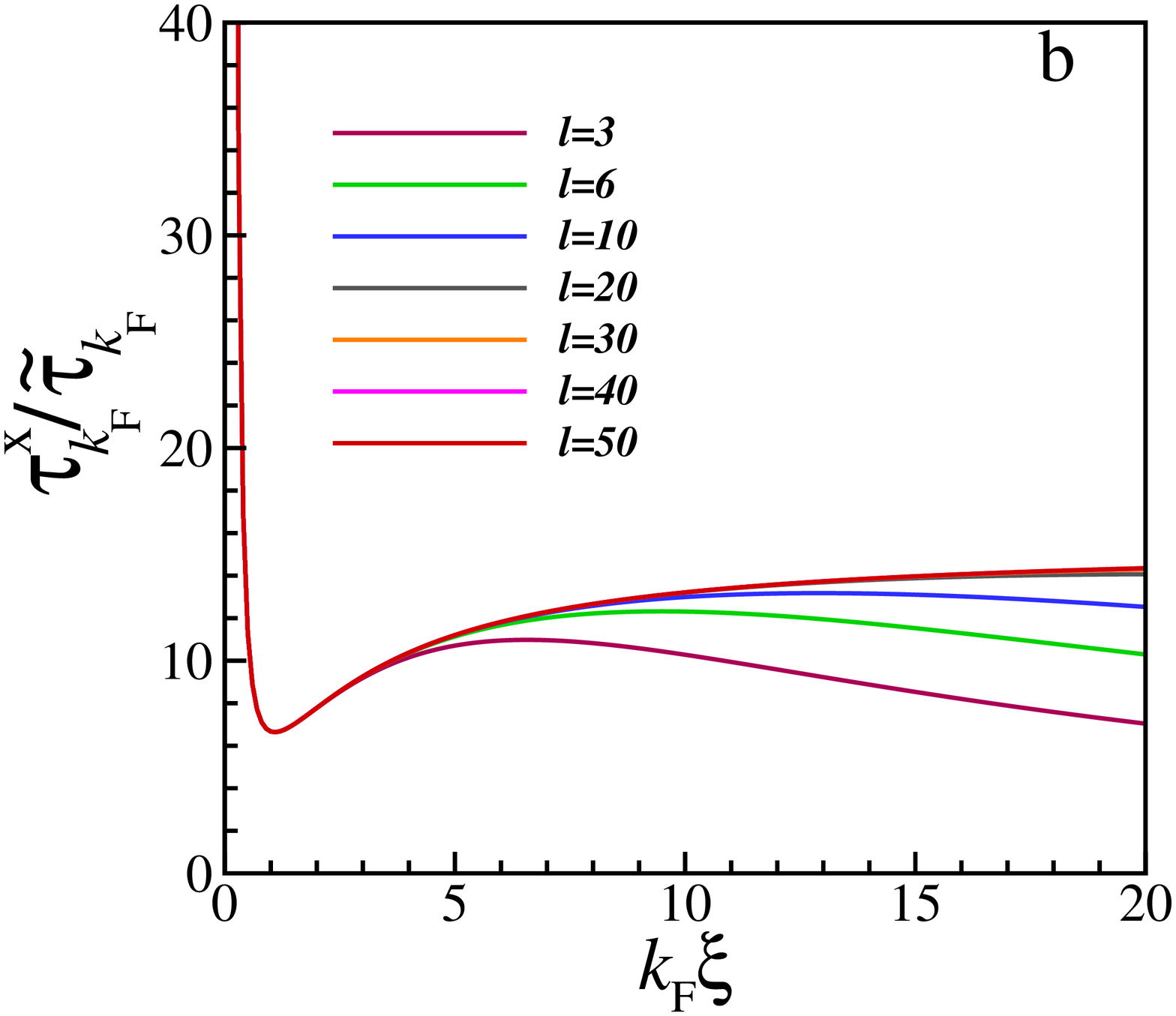}}
\centerline{\includegraphics[width=60mm]{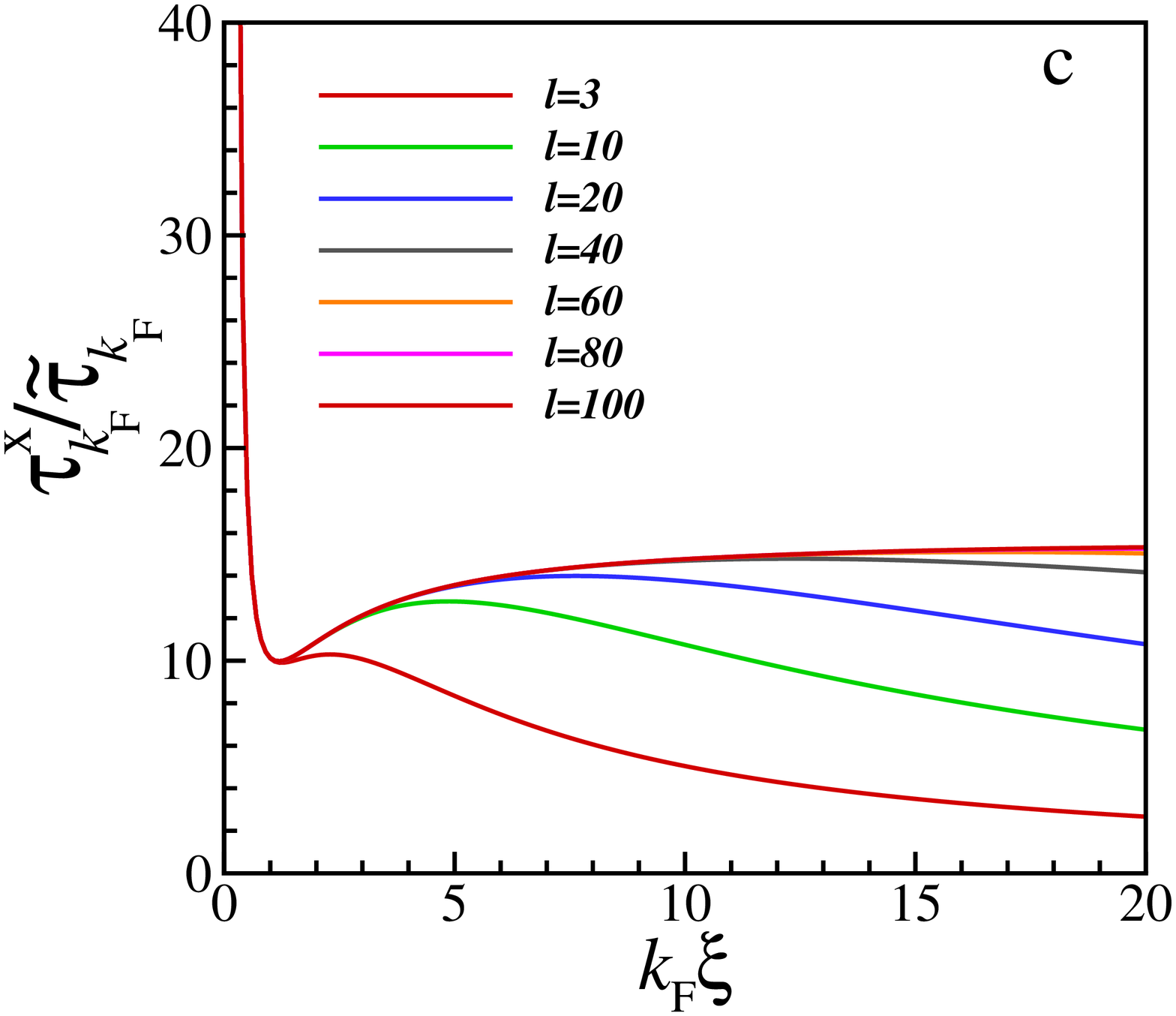}
\includegraphics[width=60mm]{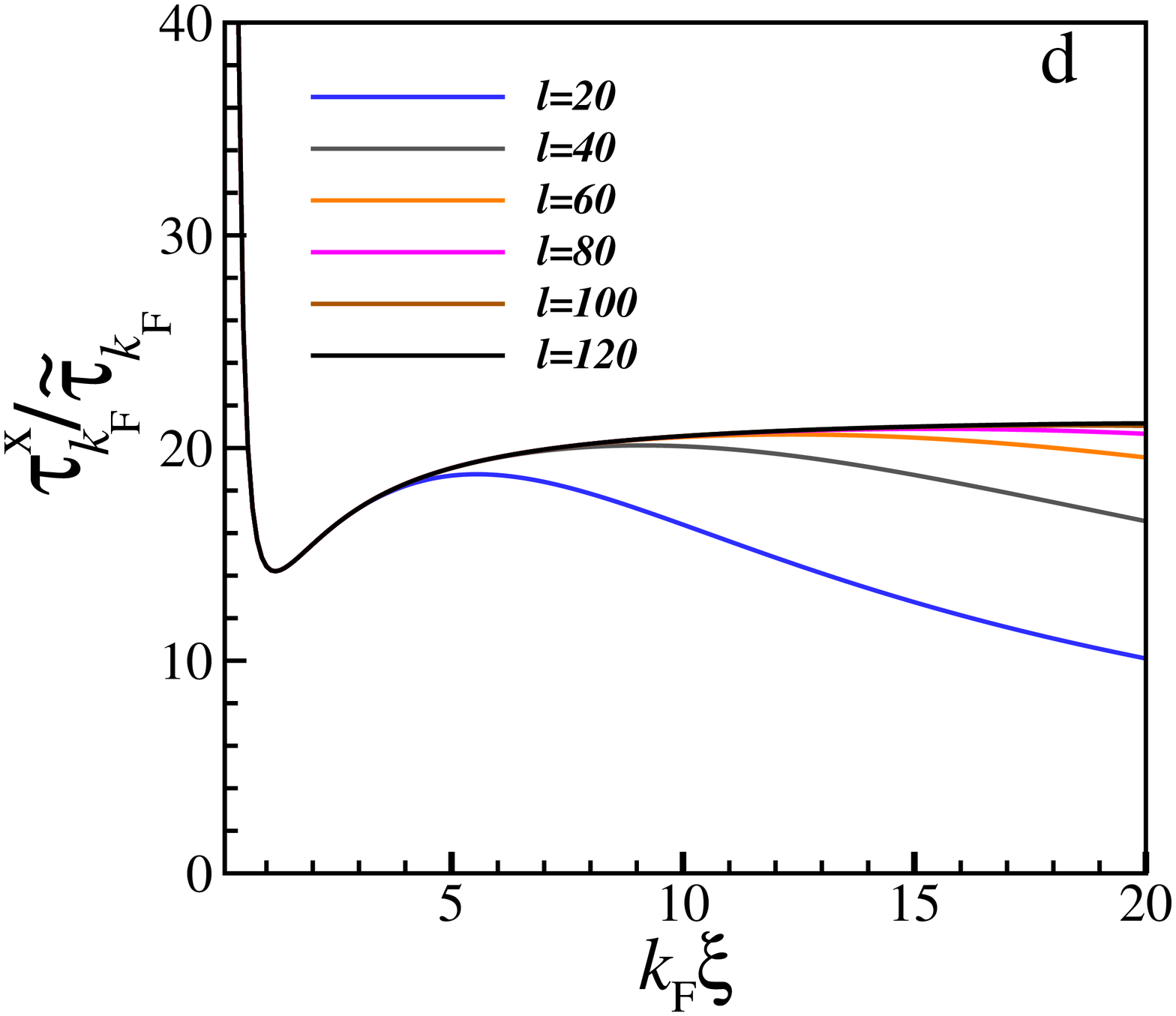}}
\caption{(Color online) The dimensionless relaxation time
$\tau^x_{k_\f}/\tilde{\tau}_{k_\f}$ versus $k_\f\xi$, for different
values of $l$ at a) $\theta=10^\circ$, b) $\theta=30^\circ$, c)
$\theta=60^\circ$ and d) $\theta=90^\circ$. By increasing $\theta$,
the TI becomes more anisotropic and we need more independent
functions to reach to a precise closed form expression for
$\tau^x_k$.} \label{fig:recursion}
\end{figure}

As we have explained in section \ref{sec:conductivity}, $a_1^c$ is
indeed the effective relaxation time $\tau_k^x$, appearing in the
relation of the surface conductivity $\sigma_{xx}$. We have plotted
in Fig. \ref{fig:recursion}, the dimensionless relaxation time
$\tau^x_{k_\f}/\tilde{\tau}_{k_\f}$ versus $k_\f\xi$, for different
values of $l$ and for tilt angles $\theta=10^\circ, 30^\circ,
60^\circ$ and $90^\circ$. Independent of $\theta$, similar to the
parameter $\lambda_0^+$, for $k_\f\xi<1$, the relaxation time
$\tau^x_{k_\f}$ does not depend on $l$ and employing a few number of
independent functions gives the precise value of $\tau^x_{k_\f}$.
For $k_\f\xi>1$, effects of anisotropy emerge and depending on the
tilt angle $\theta$, a large $l$ is required to reach a precise
closed form expression for $\tau^x_k$. When the tilt angle increases
from 0 to $90^\circ$, the TI becomes more anisotropic and the
relaxation time versus $l$ saturates at larger values of $l$. For
example, at $\theta=10^\circ$, for $k_\f\xi\gg 1$, the relaxation
time $\tau^x_k$ saturates at $l\sim 15$, however at
$\theta=90^\circ$, we need to employ $l\sim 120$ independent
functions, to attain a precise expression for $\tau^x_k$.

The other coefficients $a_{2n+1}^c$ can also be obtained in terms of
$a_1^c$, by using the recursive relations in
(\ref{eq:recursive-odd-ac}).

The number of Dirac electrons should be conserved in scattering.
This implies that the coefficient $a^c_0$ and consequently all
$a^c_{2n}$ should be zero. Furthermore, since the linear equations
obtained from the recursive relations in
(\ref{eq:recursive-odd-even-as}) are homogeneous, all the
coefficients $a_n^s$ with odd and even indices are zero. The
function $a(\phi)$ is finally given by
\begin{align}
a(\phi)=\sum_{m=0}^l
a^c_{2m+1}\cos(2m+1)\phi.\label{eq:truncation-a}
\end{align}

By analogy with the derivation of $a(\phi)$, we can show that the
recursive relations in (\ref{eq:recursive-odd-bc}),
(\ref{eq:recursive-even-bc}) and (\ref{eq:recursive-odd-even-bs})
lead to the coefficient $b_1^s$:
\begin{equation}
     \begin{aligned}
            b^{s}_{1}=\frac{W_{0}}{K_{0}-L_{0}-\lambda_{1}}=\lambda_{0}^{-}W_{0},
      \label{bs}
        \end{aligned}
     \end{equation}
where $\lambda_{0}^{-}=(K_{0}-L_{0}-\lambda_{1})^{-1}$. We have
plotted in Fig. \ref{fig:denominator}-b, the function $\lambda_0^-$
versus $\Omega$ for different values of the truncation parameter
$l$. As it is shown, for small values of $k_\f\xi$, we can obtain a
precise value for $\lambda_0^-$, by keeping only a few number of
independent functions (for example $l\simeq 3$). The other
$b^s_{2n+1}$ coefficients can be obtained by the recursive relations
in (\ref{eq:recursive-odd-even-bs}). Particle number conservation
again dictates that $b_0$ should be zero which results in
$b_{2n}^s=0$. The coefficients of the cosine functions are also zero
because the linear equations obtained from the recursive relation
(\ref{eq:recursive-odd-bc}) and (\ref{eq:recursive-even-bc}) are
homogeneous. Finally the function $b(\phi)$ is given by
\begin{align}
b(\phi)=\sum_{m=0}^l
b^s_{2m+1}\sin(2m+1)\phi.\label{eq:truncation-b}
\end{align}


\end{widetext}


\bibliography{arxiv}

\end{document}